\documentclass[11pt]{article}
\DeclareUnicodeCharacter{2212}{\ensuremath{-}}
\pdfoutput=1
\usepackage{graphicx}
\graphicspath{{./figures/}}
\usepackage{appendix}
\usepackage{braket}
\usepackage{comment}
\usepackage{fontawesome5} % For GitHub icon
\usepackage{latexsym,amsmath,amsfonts,amssymb,booktabs}
\usepackage[font=small]{caption}
\usepackage{slashed,upgreek,amscd,cancel,tensor,color}
\usepackage{adjustbox}
\usepackage{soul}
\usepackage{multirow}
\usepackage[normalem]{ulem}
\usepackage[numbers,compress,square]{natbib}
\usepackage{epsfig,latexsym}
\usepackage{amsmath}
\usepackage[pdfencoding=auto]{hyperref} 
\usepackage{url}
\numberwithin{equation}{section}
\usepackage{doi}
\usepackage{subcaption}
\definecolor{MyBlue}{rgb}{0.15,0.15,0.70}
\definecolor{linkblue}{rgb}{0,0,0.8}
\definecolor{linkgreen}{rgb}{0,0.5,0}

\hypersetup{
colorlinks=true,
citecolor=linkgreen,
linkcolor=linkblue,
urlcolor=linkblue
}

\input epsf
\usepackage{amssymb}
\usepackage{amsmath}
\usepackage{amsfonts}
\usepackage{upgreek}
\usepackage{latexsym}
\usepackage{stfloats}
\usepackage{afterpage}

\newcommand{\hinvMpc}{\,h\, {\rm Mpc}^{-1}\,}
\newcommand{\Mpcinvh}{\, {\rm Mpc} \,h^{-1}\,}

\usepackage{xspace}
\newcommand{\Python}{\texttt{Python}\xspace}
\newcommand{\JAX}{\texttt{JAX}\xspace}
\newcommand{\PyBird}{\texttt{PyBird}\xspace}
\newcommand{\JAXBird}{\texttt{PyBird-JAX}\xspace}
\newcommand{\JAXEmu}{\texttt{PyBird-Emu}\xspace}
\newcommand{\Taylor}{\texttt{PyBird-Taylor}\xspace}

\begin{document}
\def\thefootnote{\fnsymbol{footnote}}
%\begin{titlepage}
\setcounter{page}{1} \baselineskip=15.5pt \thispagestyle{empty}

\vspace*{-25mm}

\vspace{0.5cm}

\begin{center}

{\Large \bf \texttt{PyBird-JAX}: Accelerated inference in large-scale structure with\\[0.2cm] 
model-independent emulation of one-loop galaxy power spectra
}  \\[0.7cm]
{\large   Alexander Reeves${}^{1}$,  Pierre Zhang${}^{1,2,3}$, and Henry Zheng${}^{4,5}$\\[0.7cm]}

\end{center}

\begin{center}

\vspace{.0cm}

\begin{small}

{ \textit{  $^{1}$ Institute for Particle Physics and Astrophysics, ETH Z\"urich, 8093 Z\"urich, Switzerland}}
\vspace{.05in}

{ \textit{  $^{2}$ Institut f\"ur Theoretische Physik, ETH Z\"urich, 8093 Z\"urich, Switzerland}}
\vspace{.05in}

{ \textit{ $^{3}$ Dipartimento di Fisica “Aldo Pontremoli”, Universit\`a degli Studi di Milano, 20133 Milan, Italy}}

{ \textit{  $^{4}$ Stanford Institute for Theoretical Physics, Physics Department, Stanford University, Stanford, CA 94306}}
\vspace{.05in}

{ \textit{  $^{5}$ Kavli Institute for Particle Astrophysics and Cosmology, SLAC and Stanford University, Menlo Park, CA 94025}}
\vspace{.05in}

\end{small}
\end{center}

\vspace{0.5cm}

\begin{abstract}
We present \texttt{PyBird-JAX}, a differentiable, \texttt{JAX}-based implementation of \PyBird, using internal neural network emulators to accelerate computationally costly operations for rapid large-scale structure (LSS) analysis. 
\texttt{PyBird-JAX} computes one-loop EFTofLSS predictions for redshift-space galaxy power spectrum multipoles in 1.2 ms on a CPU and 0.2 ms on a GPU, achieving 3-4 orders of magnitude speed-up over \texttt{PyBird}.
The emulators take a compact spline-based representation of the input linear power spectrum $P(k)$ as feature vectors, making the approach applicable to a wide range of cosmological models. 
We rigorously validate its accuracy against large-volume simulations and on BOSS data, including cosmologies not explicitly represented in the training set. 
Leveraging automatic differentiation, \texttt{PyBird-JAX} supports Fisher forecasting, Taylor expansion of model predictions, and gradient-based searches. 
Interfaced with a variety of samplers and Boltzmann solvers, \texttt{PyBird-JAX} provides a high-performance, end-to-end inference pipeline. 
Combined with a symbolic-$P(k)$ generator, a typical Stage-4 LSS MCMC converges in minutes on a GPU. 
Our results demonstrate that \texttt{PyBird-JAX} delivers the precision and speed required for upcoming LSS surveys, opening the door to accelerated cosmological inference with minimal accuracy loss and no pretraining. 
In a companion paper~\cite{paper2}, we put \texttt{PyBird-JAX} to use in achieving LSS marginalised constraints free from volume projection effects through non-flat measures. 
\end{abstract}

\newpage

\tableofcontents

\vspace{.5cm}

\def\thefootnote{\arabic{footnote}}
\setcounter{footnote}{0}

%%%%%%%%%%%%%%%%%%
%
%
%
%

\section{Introduction}  \label{sec:intro}

Cosmology has entered an era of precision science. The increasing data volume gathered by wide-field instruments demands better accuracy in the modeling of the large-scale structure (LSS). 
In order to access information residing beyond linear scales, it is necessary to introduce progressively more nuisance parameters to marginalise over our ignorance of nonlinearities and astrophysical processes. At the same time, the multitude and diversity of cosmological probes greatly increase the dimension of the parameter space to explore when performing combined probes analyses.
In the context of spectroscopic experiments, full-shape analyses introduce $\sim \mathcal{O}(10-40)$ nuisance parameters for the power spectrum or bispectrum modeled at the one-loop level from the Effective Field Theory of Large-Scale Structure (EFTofLSS)~\cite{Desjacques:2016bnm,Perko:2016puo,DAmico:2022ukl,Philcox:2022frc}. 
A quick estimate for the next observables in this program, \textit{i.e.}, two-loop power spectrum and one-loop trispectrum, yields up to $\sim \mathcal{O}(100)$ nuisance parameters considering both the bias expansion~\cite{Schmidt:2020ovm,Donath:2023sav} and extra counterterms in redshift space~\cite{DAmico:2022ukl}. 
A clustering analysis with $\sim \mathcal{O}(10)$ tracers/redshift targets from \textit{e.g.}, DESI~\cite{DESI:2024aax} and Euclid~\cite{Euclid:2024yrr} naively requires scanning over $\sim \mathcal{O}(10^3)$ parameters, depending on the observables considered.  
Current $N$-point function analyses already marginalise over up to $\sim \mathcal{O}(100)$, depending on the refinements in the predictions and the setup considered (see \textit{e.g.}, refs.~\cite{DAmico:2019fhj,Ivanov:2019pdj,Chen:2021wdi,DAmico:2022osl,Philcox:2022frc,DESI:2024jis,DESI:2024hhd}). 
Exploring parameter spaces with a large number of dimensions can be prohibitively time and resource consuming with traditional sampling methods. Where possible, analytical marginalisation can significantly reduce the dimensionality of the problem~\cite{1003.1136,DAmico:2019fhj,DAmico:2022osl}. More generally, recent years have seen the advent of new computational techniques particularly well suited to tackle this problem. 
For example, emulators, designed to interpolate smooth, cosmological predictions, can significantly speed up the likelihood call with controlled accuracy loss. 
Furthermore, differentiable likelihoods enable efficient exploration of high-dimensional parameter space using gradient-informed techniques. 
Straightforward solutions that capture both of these benefits consist in Taylor-expanding the cosmological observables around a fiducial point~\cite{Cataneo:2016suz,Colas:2019ret,Lai:2024bpl}, training neural networks (NNs), Gaussian processes, or other methods, to emulate them (see \textit{e.g.}, refs.~\cite{Auld:2006pm,Albers:2019rzt,Zennaro:2021bwy,Arico:2021izc,Mootoovaloo:2021rot,SpurioMancini:2021ppk,Nygaard:2022wri,Gunther:2022pto,Bonici:2022xlo,Reeves:2023kjx,2311.15865,Zhang:2025cdv,2503.13183}), or leveraging modern coding languages such as \JAX\footnote{\url{https://docs.jax.dev/}} or \texttt{Julia}\footnote{\url{https://julialang.org/}} (see \textit{e.g.}, refs.~\cite{Campagne:2023ter,Bonici:2023xjk,Hahn:2023nvb,Ruiz-Zapatero:2023hdf,Balkenhol:2024sbv,Reeves:2025axp,Reymond:2025ixl}). 

In this paper, we present an updated version of \PyBird: the \texttt{Python} code for Biased tracers in redshift space~\cite{DAmico:2020kxu}. \PyBird provides nonlinear predictions and likelihoods for cosmological correlators based on the EFTofLSS. As one of the earliest implementations of this kind, extensively tested against multiple $N$-body simulations with diverse halo-galaxy connection models~\cite{Nishimichi:2020tvu,Zhang:2021yna,DAmico:2021ymi,DAmico:2022osl,Simon:2022csv,Zhao:2023ebp,Lai:2024bpl}, \PyBird has been benchmarked against other pipelines~\cite{Simon:2022lde}, particularly within the contexts of DESI~\cite{Lai:2024bpl,Maus:2024sbb} and Euclid~\cite{Euclid:2023bgs}. 
\PyBird has been used to put constraints on a wide variety of physics (see \textit{e.g.},~refs.~\cite{DAmico:2020kxu,Bottaro:2023wkd,Poudou:2025qcx,Calderon:2025xod,DAmico:2022gki,Piga:2022mge,Spaar:2023his,Taule:2024bot,Lu:2025gki,Glanville:2022xes,Lague:2021frh,Gonzalez:2020fdy,Allali:2021azp,Allali:2023zbi,Simon:2022ftd,Niedermann:2020qbw,Cruz:2023cxy,DAmico:2020ods,Smith:2020rxx,Simon:2022adh,Gsponer:2023wpm}) from
various LSS datasets~\cite{DAmico:2022osl,Simon:2022csv,Zhao:2023ebp}, including the first results from the full shape of DESI first data release~\cite{DESI:2024jis,DESI:2024hhd}. 
A number of alternative codes for computing the one-loop power spectrum have also been developed~\cite{Chudaykin:2020aoj,Chen:2020fxs,Eggemeier:2022anw,Carrilho:2022mon,Noriega:2022nhf,2503.16160,Linde:2024uzr}, further reflecting the community’s active interest in this area. 
While these codes agree in their core calculations, each offers distinct features and implementation details. We encourage readers to explore these complementary tools.

In this work, we revisit the implementation of \PyBird in several ways, resulting in a fully differentiable pipeline with significantly boosted performance. While keeping compatibility with most cosmological codes, Boltzmann solvers, numerical samplers, etc., within the rich \texttt{Python} environment, we upgrade \PyBird to \texttt{JAX}. Our new \texttt{JAX} implementation, that we dub \JAXBird, allows for just-in-time compilation to dedicated computing hardware, vectorisation, and automatic differentiation (AD). Furthermore, we accelerate the slowest parts of \JAXBird by emulating their computation with NN-based emulators. We will refer to the latter scenario as \JAXEmu.   
Leveraging the fact that \PyBird takes as input the linear matter power spectrum which can be readily decomposed onto a compact set of basis functions, we train NNs only on the decomposition coefficients, rendering our emulator explicitly \emph{cosmology‐independent}.  This strategy contrasts with existing \PyBird\ emulators and similar codes (see, \textit{e.g.},~\cite{DeRose:2021pqx,Donald-McCann:2022pac,Ramirez:2023ads,Lai:2023nil,Trusov:2024mmw,Bonici:2025ltp}), which require retraining for every cosmological model and usually demand additional accuracy checks against full, non-emulated analyses (see however ref.~\cite{2503.13183} for an on-the-fly efficient emulation strategy during sampling, though applied on different observables). At present, the emulation is carried on the redshift-space one-loop galaxy power spectrum, however our method is general and can apply to other observables such as higher-loop corrections and higher-$N$ point functions, as well as multi-tracer statistics. 
Overall, our improvements result in massive acceleration approximately three (four) orders of magnitude on CPU (GPU) units, making one evaluation with \JAXEmu as fast as 1.2 ms (0.2 ms). 
Numerical accuracy on predictions are found below $0.1\sigma$ relative to a representative Stage-4 spectroscopic survey volume of $V_{\rm eff} = 50\,\textrm{Gpc}^3$, for 95\% of the distribution of an independent validation set. 
This makes \JAXBird sufficiently accurate for galaxy clustering analyses in upcoming surveys over the next decade.
\JAXBird is publicly released in the same repository of \PyBird.\footnote{See \url{https://github.com/pierrexyz/pybird} and \url{https://pierrexyz.github.io/pybird/}}  
\JAXBird can be activated via a simple backend switch on top of \PyBird as core implementation are shared, with the difference being the underlying libraries—\JAX or \Python—called during execution.

Our paper is organised as follows. 
In sec.~\ref{sec:emulation}, we present a general, emulator-based, model-independent method for LSS observables, which we implement to accelerate the evaluation of \PyBird one-loop galaxy power spectra.  
Sec.~\ref{sec:interface} summarises the new features introduced in \JAXBird, with a focus on AD and enhanced compatibility with various Boltzmann solvers and samplers.  
In sec.~\ref{sec:applications}, we benchmark the fully upgraded pipeline on challenging cosmological analyses, including simulations, BOSS data, and a Stage-4 LSS spectroscopic mock survey.  
We conclude in sec.~\ref{sec:conclusions}.  
Additional technical details on the implementation of IR-resummation in \PyBird are provided in app.~\ref{app:resum}, while further accuracy assessments of the emulator are presented in app.~\ref{app:emuacc}.

As an important application, our companion paper~\cite{paper2} leverages the differentiability of \JAXBird to construct robust estimators for parameter inference. We show that, with properly defined expectation values, the posterior means of the parameters of interest can be accurately recovered—without significant bias from volume projection effects—even when marginalizing over a large number of nuisance parameters.  
By offering practical tools and theoretical insights in parameter inference, this two-paper series enables efficient and robust cosmological analysis from the LSS in the new era of precision cosmology. 

%%%%%%%%%%%%%%

\section{Model-independent emulation of one-loop galaxy power spectra} \label{sec:emulation}
In this section, we present our methodology for implementing model-agnostic emulators in \PyBird. The most expensive steps of the evaluation of a given galaxy power spectrum using \PyBird are the loop integrals and the IR-resummation as described in ref.~\cite{DAmico:2020ods} (see also ref.~\cite{Zhang:2021yna}). As we show in sec.~\ref{sec:performance}, whilst implementing these routines in \texttt{JAX} and using just-in-time compilation already gives a speed-up of a factor of $\mathcal{O}(10)$, we can gain a further speed boost by a factor of $\mathcal{O}(100)$ by emulating these components with \texttt{JAX}-based NNs. 
%This emulation is designed to be internal to the \PyBird code. 
%After factorising the EFT parameters and growth rates, the remaining pieces in these operations, which are the expensive steps, depend only on the input linear matter power spectrum $P_{\rm lin}(k)$. 
Schematically, the galaxy power spectrum in redshift space $P_{g,r}$ reads
\begin{equation}\label{eq:master}
P_{g,r}(k,\mu,z) = \sum_{i=0}^N \sum_{j=0}^N b_i b_j \, \mu^{r_{ij}}f(z)^{s_{ij}} P^{ij}(k,z) \ , 
\end{equation}
where $b_i$, $i=1, \dots, N$, are the EFT parameters describing the one-loop power spectrum in redshift space in the EFTofLSS (see sec.~\ref{sec:applications} for details), $b_0 \equiv 1$, $\mu$ is the cosine of angle of $\pmb k$ with the line-of-sight, and $r_{ij}, s_{ij}$ are some associated integer powers that are non-zero for line-of-sight-dependent terms arising in redshift space.  
Here $P^{ij} \equiv P^{ij}[P_{\rm lin}]$ are scale-dependent functions that generically depend on $P_{\rm lin}$, the input linear matter power spectrum, which encodes their cosmology dependence.  
In app.~\ref{app:resum}, we show how the IR-resummation scheme used in \PyBird can also be written in a functional form akin to eq.~\eqref{eq:master}.  
The most expensive pieces correspond to loop integrals of the 22-diagram type (see eq.~\eqref{eq:22}).

We train an emulator for the loop integrals and the IR-resummation and use an input parameter space which represents a decomposition of the linear matter power spectrum as described in the next section.  This approach is independent of specific cosmologies and redshifts, and can thus be used for any cosmological exploration and survey specification. This means the emulators will not have to be retrained when exploring a new specific model or survey set up. Notice that, in principle, this strategy could also be used to emulate all parts of the galaxy power spectrum. 
However, the linear galaxy power spectrum and EFT counterterms are trivially dependent on $P_{\rm lin}$ and hence do not need to be emulated as their evaluation is already rapid. 
By keeping these pieces computed using the analytic expression, we limit the impact of numerical inaccuracy in the emulation to only the loop and IR-resummation contributions. 
Since these pieces represent only a fraction of the total signal of $\sim \mathcal{O}(10\%)$, achieving $\sim 1\%$ relative accuracy in the emulation leads to a only a $\sim 0.1\%$ relative error in the total galaxy power spectrum. 
We now detail our procedure to find an optimised decomposition of the input linear power spectrum for the input to the NN emulators. 

\subsection{Model-independent NN-based emulator}

\paragraph{Optimised decomposition of linear power spectrum }
A desired parametrisation for the linear matter power spectrum has the following properties. First, it should be general enough to encompass a variety of shapes and features (arising from the diversity of cosmological considerations). Second, the number of parameters should be reasonably small enough such that the NN can accurately interpolate between across the input parameter space without requiring a prohibitively large number of training samples. 
To meet these constraints, we choose to interpolate $\log P_{\rm lin}(k)$ with a linear spline on predefined $k$-knots in $\log k$ space.  To reduce dynamic range, we normalise $P_{\rm lin}$ at its maximum to $1$ by dividing by the maximum value of the power spectrum, $P_{\rm max}$. We optimise (through Monte Carlo sampling) the knot placements on the \texttt{CosmoRef} bank detailed below and in table~\ref{tab:priors}. 
Explicitly, we minimise the residual differences between normalised linear power spectra and the corresponding recovered spline interpolation weighted by the inverse of a Gaussian covariance to mimic data sensitivity together with a $(k/k_{\rm NL})^2$ scaling as a proxy of the loop $k$-dependence and size. We use a total of $80$ $k$-knots spaced between $(k_{\rm min}, k_{\rm max}) = (0.0001, 0.7) \hinvMpc$ which represents a generous window in $k-$space over which cosmological data are sensitive.
Outside of this range, $\log P_{\rm lin}(k)$ is extrapolated linearly in $\log k$. 
Here we take advantage of the smallness of the loop and IR-resummation at low-$k$ and their relatively small sensitivity (at all $k$'s) to the IR-part of $P_{\rm lin}$. 
Furthermore, the UV-part of $P_{\rm lin}$, whilst changing the values of the loop integrals, will in fact not matter in the full prediction. 
This follows the fact that the UV-sensitivity of the loop integrals are absorbed by the counterterms, assuming that the actual values of the latter do not matter for the user. 
Would the actual values of the counterterms matter (as when one wishes to impose informative prior), \PyBird includes a numerical method to control the UV matching explicitly by computing the contributions to the counterterms from the UV limit of the loop integrals.\footnote{Explicitly, focusing on the 13-loop in real space for definiteness, the UV limit of the loop kernels $K_{13}$ expanded up to the relevant order for our computation is
\begin{equation}
\lim_{\frac{q}{k}\rightarrow\infty} K_{13}(q, k) = \alpha_0 \left( \frac{k}{q}\right)^0 + \alpha_2 \left( \frac{k}{q}\right)^2 + \dots \ ,
\end{equation}
where $\alpha_0, \alpha_2, \dots$, are rational numbers. 
\PyBird then adds, for instance, a contribution in the form of the counterterm, 
\begin{equation}
\Delta P_{13}(k) = 4\pi \, \alpha_2 k^2 P_{\rm lin}(k) \int_0^{\infty} dq\, \Delta P_{\rm lin}(q) \ ,
\end{equation}
where $\Delta P_{\rm lin}(k) = P^{\rm ref}_{\rm lin}(k)-P_{\rm lin}(k)$, with $P_{\rm lin}(k) \approx P^{\rm ref}_{\rm lin}(k)$ for $k \leq k_{\rm max} = 0.7 \, \hinvMpc$. 
Consisting in a 1D integral of the UV limit of the loop kernel times the difference between the log-interpolated $P_{\rm lin}$ with a specified reference $P^{\rm ref}_{\rm lin}$, this evaluation is relatively inexpensive. }

\begin{table}[h!]
    \centering
      \vspace{3pt}
  \begin{tabular}{@{}l c@{}}
    \toprule
    Parameter & Prior (Gaussian $\times$ flat bound) \\
    \midrule
    $\ln(10^{10} A_s)$         & $\mathcal{N}(3.04,\,0.5) \,\times\, [2,\,4]$          \\
    $h$                         & $\mathcal{N}(0.68,\,0.10) \,\times\, [0.40,\,1.10]$   \\
    $\Omega_m$                  & $\mathcal{N}(0.310,\,0.05) \,\times\, [0.20,\,0.40]$  \\
    $n_s$                       & $\mathcal{N}(0.965,\,0.10) \,\times\, [0.70,\,1.20]$  \\
    $\omega_{\rm b}$                  & $\mathcal{N}(0.0223,\,0.01) \,\times\, [0.010,\,0.030]$\\
    $\sum m_\nu\,[\mathrm{eV}]$ & $\mathcal{N}(0.10,\,0.50) \,\times\, [0,\,1.5]$       \\
    $\Delta N_\mathrm{eff}$      & $\mathcal{N}(0,\,1.0) \,\times\, [-2,\,2]$            \\
    $\Omega_k$                  & $\mathcal{N}(0,\,0.10) \,\times\, [-0.3,\,0.3]$       \\
    $w_0$                       & $\mathcal{N}(-1.0,\,0.50) \,\times\, [-2,\,0]$        \\
    $w_a$                       & $\mathcal{N}(0,\,1.0) \,\times\, [-2,\,2]$            \\
    $z$                         & $\operatorname{LogNormal}(0.8,\,0.14) \,\times\, [0.005,\,4.0]$ \\
    \bottomrule
  \end{tabular}
\caption{\textbf{The \texttt{CosmoRef} bank} --- Prior ranges used to construct a bank of linear matter power spectra guiding the design of the input coverage for our model-independent emulator. 
Cosmological parameters are sampled from wide Gaussian distributions centred on values close to the \textit{Planck} $\Lambda$CDM preferred values (and nominal standard values for beyond-$\Lambda$CDM parameters). 
The Gaussian widths are set to approximately two to three times the $1\sigma$ uncertainties from the BOSS full-shape mock analysis described in sec.~\ref{sec:BOSS}, with the exception of $h$, for which we consider a larger width given current debate around its value. 
For the baryon abundance $\omega_{\rm b}$, we consider a width of $\sim 25$ times the BBN prior instead. 
Gaussians are truncated at roughly $5\sigma$ or more relative to BOSS errors. 
Redshifts $z$ are sampled from the clipped log-normal distribution described in the text.
The \texttt{CosmoRef} bank used to design the Gaussian copula distribution over the emulator’s input space comprises $10^6$ linear matter power spectrum computed via the Boltzmann solver \texttt{CLASS}, drawn from the above prior within a latin hypercube. 
}
\label{tab:validation_range}
\end{table}

\begin{figure}[!ht]
    \centering
    \includegraphics[width=0.99\textwidth]{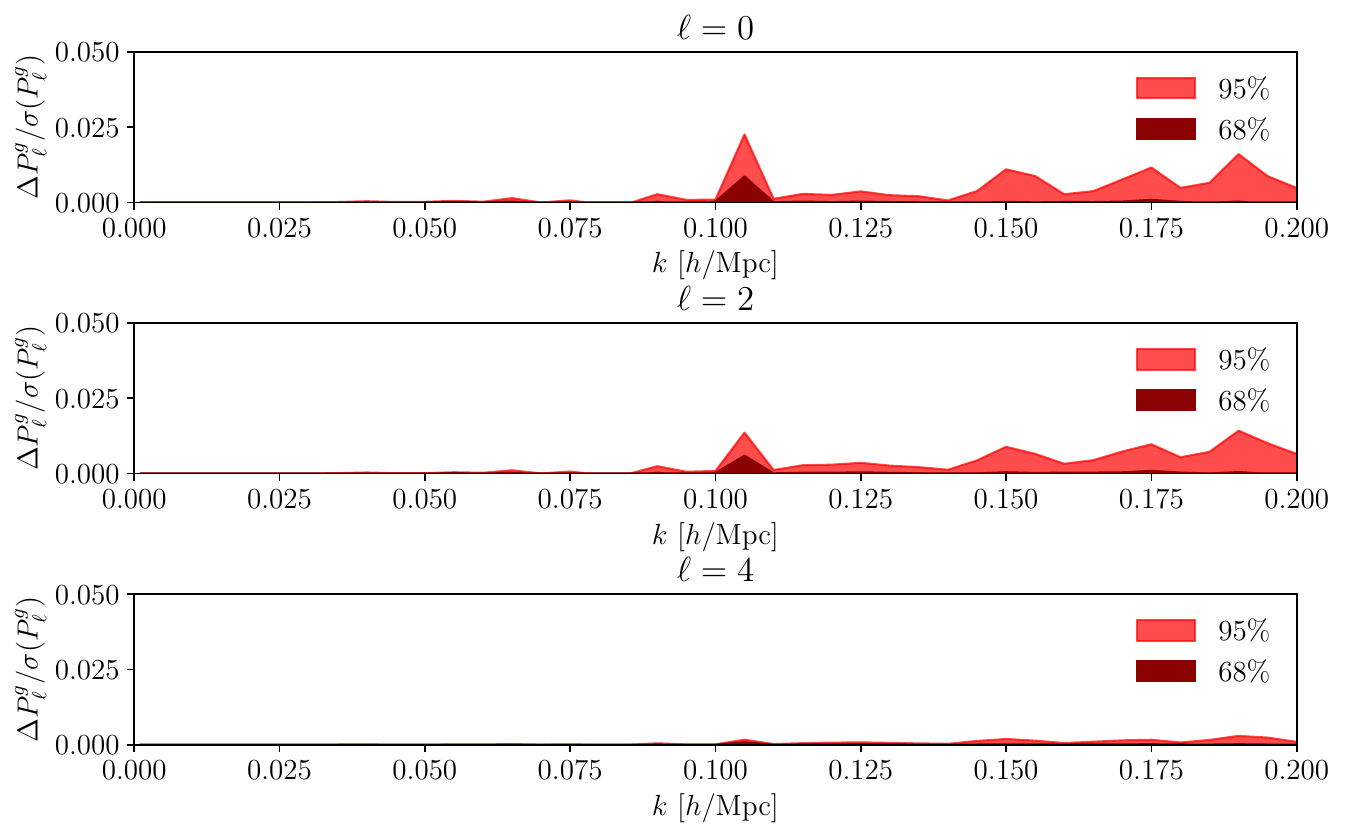}
    \caption{\textbf{Spline decomposition accuracy} --- Distribution of errors in the galaxy power spectrum multipoles computed with \PyBird, comparing results obtained using either the input linear matter power spectrum or its spline-reconstructed counterpart, across the \texttt{CosmoRef} testing bank described in table~\ref{tab:validation_range}. 
    Errors are shown relative to representative uncertainties expected for Stage-4 LSS surveys. }
    \label{fig:spline_residue}
\end{figure}

To test the spline decomposition we use a bank of $10^5$ linear matter power spectra in cosmologies drawn from a 11-dimensional cosmological latin hypercube, dubbed the \texttt{CosmoRef} bank. 
One dimension consists in a truncated log-normal draw for the redshifts which is clipped between $z\in [0.005,4]$ and the log-normal distribution is chosen to peak at $z=0.6$. This choice is motivated by the fact that the linear matter power spectrum scales as
\(P_{\mathrm{lin}}(k,z)\propto D^{2}(z)\simeq (1 \times z)^{-2}\); 
a \emph{uniform} draw in \(z\) therefore yields a distribution of amplitudes that is very concentrated near zero. 
The log-normal distribution in $z$ shifts the probability mass away from $0$ while retaining a rather flat high-$z$ tail, producing a flatter distribution in amplitude (and in growth rate $f$ following the same argument) more suitable for agnostic training. The remaining dimensions consist of ten cosmological parameters: $\{\omega_{\rm b}, \Omega_m, h, n_s, \ln(10^{10} A_s), \Omega_k, \sum m_\nu, N_{\rm eff}, w_0, w_a\}$, corresponding respectively to the physical baryons abundance, the fractional dark matter abundance, the reduced Hubble constant, the spectral tilt, the log of the rescaled amplitude of the primordial power spectrum, the fractional curvature abundance, the total neutrino mass, the relativistic number of species, and the dark energy equation of state parameters. Each cosmological parameter is distributed in the hypercube according to a Gaussian centered on values close to \textit{Planck} preferred ones~\cite{Planck:2018vyg}, with standard deviation of about two or three times the size of the $68\%$CL uncertainties $\sigma$ obtained on BOSS~\cite{DAmico:2022osl}, and clipped at roughly $5\sigma$. 
The parameter distributions are shown in table~\ref{tab:validation_range}. 

Figure~\ref{fig:spline_residue} presents the quantiles of the residuals distribution on the galaxy power spectrum multipoles computed with \PyBird taking either as input the full linear matter power spectrum or its spline over the $80$ optimised $k$-knots. Each residual is weighted by the forecast $1\sigma$ uncertainties of a representative Stage-4 LSS survey (as described in sec.~\ref{sec:performance}), and the statistic is evaluated across the entire test bank. 
With $N_k = 80$ knots and optimised placements, we find that the $95\%$ quantile of the error distribution does not reach greater than 3\% of Stage-4 LSS uncertainties.  
The error from the spline decomposition is thus virtually negligible given our target precision. 

In principle, other decompositions, potentially more suitable to capture the features in $P_{\rm lin}$ with a reduced number of basis functions, could be considered
(see \textit{e.g.}, refs.~\cite{Simonovic:2017mhp,Anastasiou:2022udy,Bakx:2024zgu}). 
However, these typically entail tradeoffs in flexibility and coverage, making the emulator less general to input cosmologies (and thus more prone to out-of-distribution anomalies) or leading to a loss of interpretability. 
For instance, the \texttt{FFTLog} decomposition used in refs.~\cite{Simonovic:2017mhp, Anastasiou:2022udy} provides a useful basis in which to compute loop integrals analytically. 
However, as a complex power-law decomposition, the broadband of the linear matter power spectrum typically requires many basis functions to be described to sufficient accuracy. 
For reference, \texttt{PyBird} uses $N_{\rm FFT}=512$, much larger than $N_{\rm knots}=80$. 
In contrast, the \texttt{COBRA} method introduced in ref.~\cite{Bakx:2024zgu} allows for an optimal decomposition with very few basis functions, $N \sim 10$. 
However, this is optimised for a given cosmological training bank and would require re-testing when applied to a model outside of the original training bank.
Meanwhile, with our simple-minded choice, the spacing between $k$-knots is on average about $\Delta k \sim 0.01$, which is enough to resolve to good accuracy the features in $P_{\rm lin}(k)$ from most cosmologies. 
If one doubts the accuracy of the parametrisation \textit{e.g.}, when exploring fast oscillating features with period below this $k$-resolution, the emulator can be simply switched off whilst still maintaining the speed boost of the pure \texttt{JAX} implementation.

\paragraph{Input emulator parameter space } 
Our input parameter space thus mainly consists in the log-power band values at the $k$-knots  of the spline previously chosen. 
Note that the normalisation factor $P_{\rm max}$ does not need to enter the emulator for the loop integrals, as the loop integrals will be simply rescaled by $P_{\rm max}^2$. 
The same argument applies to the correction terms appearing in the IR-resummation scheme reviewed in app.~\ref{app:resum}. 
As shown by eq.~\eqref{eq:resum_master}, they would also be rescaled by powers of $P_{\rm max}$. 
However, for practical reasons related to memory usage, we use another emulator depending further on $P_{\rm max}$ and the growth factor $f$ for the IR-resummation pieces. 
The inclusion of $f$ allows us to compute the matrix coefficients $\mathcal{Q}^{\ell\ell',m,\alpha}||_{N-j}(f)$ in eq.~\eqref{eq:resum_master}, such that we save only one IR-correcting term per linear / loop / counter- term, instead of $N_\ell \times M \times N_\alpha \sim 200$. 
We find that emulating the full IR-resumed loop pieces, as opposed to a separate loop and IR-correction piece, yields the best performance in both speed and accuracy. As mentioned before, we use the full analytic expressions for the linear and counterterm pieces so for these we train only the corresponding IR-correction parts. The next section details our prior choice for the input parameters, selected to ensure wide coverage for different cosmological models.

\begin{figure}[!ht]
    \centering
    \includegraphics[width=0.8\textwidth]{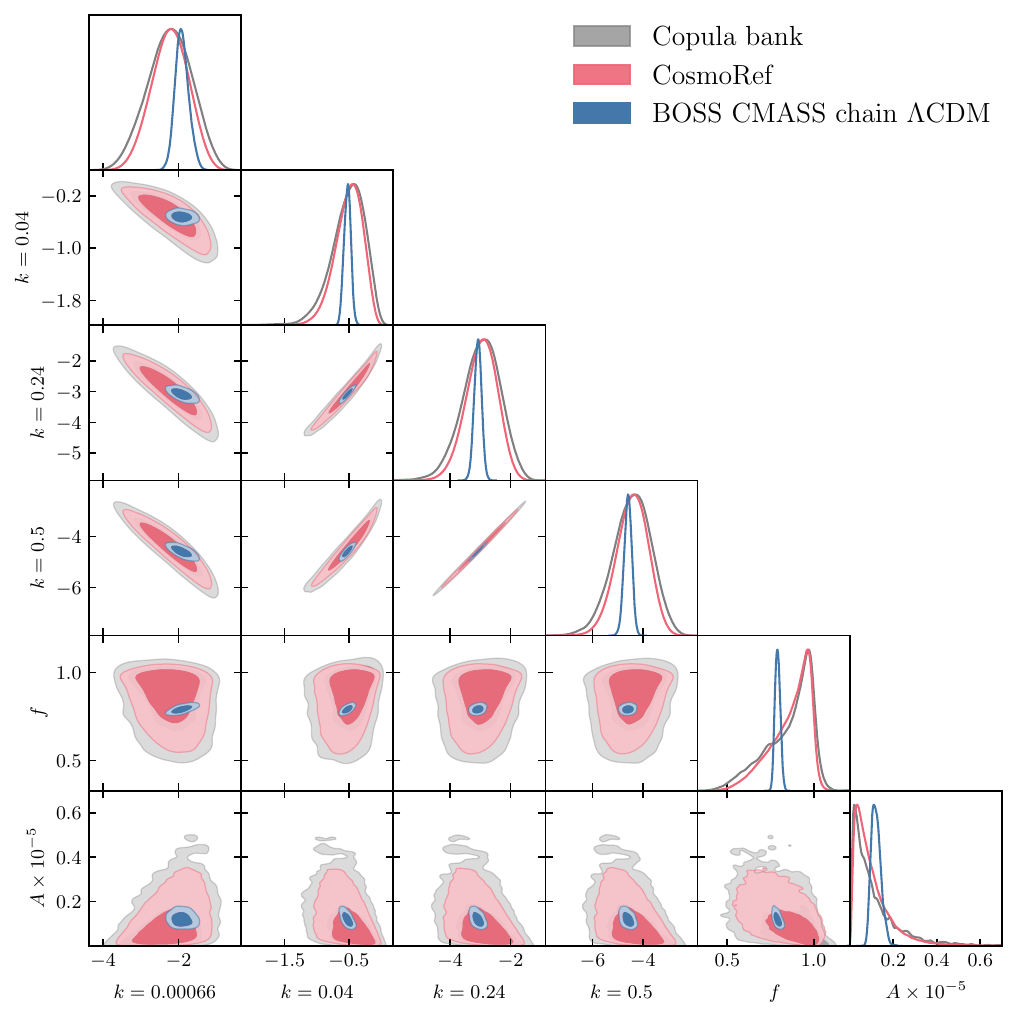}
    \caption{\textbf{Representative coverage over emulator input parameter space} --- Inflated Gaussian copula distribution over the emulator training input space, shown for selected knots (all ks in $\hinvMpc$), maximal power amplitude $A = P_{\mathrm{max}}$ (in $(\Mpcinvh)^3$), and growth factor $f$. The original reference \texttt{CosmoRef} bank and samples from BOSS CMASS $\Lambda$CDM analysis are also shown for comparison.}
    \label{fig:input_coverage}
\end{figure}

To provide the emulator with a large yet physically faithful training bank, we begin with a parent, representative ensemble, that we take to be the \texttt{CosmoRef} bank. 
This consists in \(N_{\mathrm{parent}}=10^{6}\) linear power spectra drawn according to the distribution specified in table~\ref{tab:validation_range}. 
The purpose of this bank is to inform the distribution in log-power band space to ensure some level of regularity and smoothness in the resulting representation of the linear power spectra. 
We then use a Gaussian–copula resampling strategy to generate additional, out-of-sample spectra, extrapolating beyond the restrictions from the inherent (cosmology-dependent) choice of the parent ensemble. 
Our approach is as follows. Each spectrum and their corresponding growth rate $f$ are first compressed to the 82-component feature vector
\begin{equation}
\mathbf x=\bigl(a_1,\ldots,a_{80},A,f\bigr),
\qquad 
P_{\rm max}\equiv\max_k P_{\rm lin}(k),
\qquad 
f\equiv\frac{\mathrm d\!\ln D}{\mathrm d\!\ln a},
\label{eq:xvector}
\end{equation}
where $a_j = \log P_{\rm lin}(k_j)/P_{\rm max}$, $j=1, \dots, 80$, are the logarithmic band power amplitudes on our previously chosen fixed knots $k_j$ for the input linear power spectrum spline. 
Because we desire the samples for the training distribution to respect both the empirical band power 1D marginals and the inter-knot correlations of the parent set, we decouple these two pieces with the copula construction. First, every coordinate $x_i$ is ``uniformised'' by its empirical 1D marginal cumulative distribution function (CDF) $F_i$,
\begin{equation}
u_i = F_i(x_i),
\label{eq:uniform}
\end{equation}
so that the variables \(\{u_i\}\) are individually uniform on $[0,1]$.  
We then use a copula to encode the correlations within the uniformised parameter space. Assuming a \emph{Gaussian copula} means imposing a latent normal representation
\begin{equation}
\mathbf q = \Phi^{-1}(\mathbf u)
           \;\sim\; \mathcal N(0,\Sigma),
\label{eq:latent}
\end{equation}
where $\mathbf q$ denotes the ``Gaussianised'' latent variables obtained by mapping each uniform variate $u_i$ through the probit $\Phi^{-1}$, \textit{i.e.}, through the inverse CDF of a standard normal. This transformation places the data in the space of a Gaussian copula: the multidimensional dependence is now encoded entirely in the covariance matrix $\Sigma$, which we estimate from the $10^6$ linear bank realizations.  Once $\Sigma$ is known we can generate new correlated knot amplitudes in two computationally cheap steps: draw a single sample from $\mathcal N(0,\Sigma)$ and then apply, in reverse, the sequence $\Phi$ followed by each empirical inverse CDF $F^{-1}_{i}$. The resulting mock vectors faithfully reproduce both the 1D marginals and the inter-knot correlations of the training data, yet remain more agnostic to any particular cosmological model that the bank samples themselves; they therefore provide a fast, general-purpose source of realistic spline-space samples for emulator training. 

To make the training data wider we multiply $\Sigma$ by a factor $\alpha>1$,
\begin{equation}
\tilde \Sigma = \alpha\,\Sigma  \ ,
\label{eq:siginfl}
\end{equation}
which expands every 
latent variable standard deviation by \(\sqrt\alpha\).  
New latent vectors $\tilde{\mathbf{q}}\sim \mathcal{N}(0, \tilde \Sigma)$ are mapped back through the inverse transforms \(\tilde{\mathbf u}=\Phi(\tilde{\mathbf q})\) and \(\tilde{\mathbf x}=F^{-1}(\tilde{\mathbf u})\). 
Because the rescaled latent variable 1D marginals are broader than standard normals, the \(\tilde{u}_i\) are no longer strictly uniform; instead they are skewed toward the boundaries \(0\) and \(1\). Consequently the inverse–CDF transform drives the features \(\tilde{x}_i\) deeper into the empirical tails, allowing the training bank to explore regions that lie beyond the original \texttt{CosmoRef} ensemble and to exhibit reduced inter-knot correlations. In practice we find a modest value of $\alpha=1.2$ to be a good balance between preserving the cosmologically-relevant inter-knot correlations whilst allowing for a widened parameter space coverage. The resulting bank captures the correct correlation structure between the knots to ensure physically viable power spectra (implying some level of regularity and smoothness in $k$) whilst the Gaussian copula with inflation by $\alpha$ allows for a wider class of models to be represented than in the original \texttt{CosmoRef} bank (both in the allowed amplitudes and shapes of linear matter power spectra). 
The training distribution over the emulator input power space is displayed in fig.~\ref{fig:input_coverage}. 
We note that there could be alternative ways to generalise representative parent ensemble to a model-agnostic distribution, such as training normalising flows to a latent space. 
We leave such exploration for future work.

\paragraph{Training} 
We run \JAXBird for $10^7$ samples of the feature vector $x$ defined in eq.~\ref{eq:xvector} drawn from the widened Gaussian Copula distribution over the emulator input parameter space to construct the training set. 
In practice, we emulate only the components of \PyBird that are numerically expensive, which consist in
\begin{itemize}
  \item The one-loop terms, either resummed or non-resummed, with one NN per multipole (monopole, quadrupole, hexadecapole), and one set of NNs for the resummed or non-resummed case;
  \item The IR-correction pieces for resumming the linear contributions, using one NN;
  \item The IR-correction pieces for resumming the counterterms, using one NN.
\end{itemize} 
Before training we compress every output vector with a principal–component analysis (PCA), retaining \(N_{\rm PCA}=256\) modes for each NN. Before compression the output vector size is $\mathcal{O}({10^4)}$ for each of the emulators. This yields two practical advantages: (\textit{i}) the reduced dimensionality lowers GPU memory requirements, allowing faster training; (\textit{ii}) the smaller dynamic range of the compressed targets improves numerical stability and accelerates convergence. 
We adopt the NN architecture from ref.~\cite{SpurioMancini:2021ppk}\footnote{Available at \url{https://github.com/alessiospuriomancini/cosmopower}}, implemented in \texttt{TensorFlow}~\cite{1603.04467}. Each model consists of three hidden layers with 512 neurons per layer. The activation function in each layer is given by
\begin{equation}
\label{eqt:activation}
    f(\vec{x}) \;=\; \bigg(\vec{\gamma} \;+\;\big(1 + \exp(- \vec{\beta}\cdot\vec{x})\big)^{-1}\,(1-\vec{\gamma})\bigg)\,\cdot\,\vec{x},
\end{equation}
where $\vec{x}$ is the vector of inputs and the hyper-parameters $\vec{\beta}$ and $\vec{\gamma}$ are trained alongside the usual network weights. 
We use the \texttt{ADAM} optimiser~\cite{1412.6980} with default parameters, a batch size of 1024, and train each emulator for 1000 epochs (\textit{i.e.}, full passes of the entire training dataset) for each of 5 separate \texttt{ADAM} learning rates ($LR$) that are decreased in a ladder from $LR=10^{-3}$ to $LR=5\times10^{-5}$. For this optimisation, four fifths of the training set are used as a pure training set, whilst the remaining fifth is used as an internal validation set.

\subsection{Performance}\label{sec:performance}

\begin{figure}[!ht]
    \centering
      \includegraphics[width=0.65\textwidth]{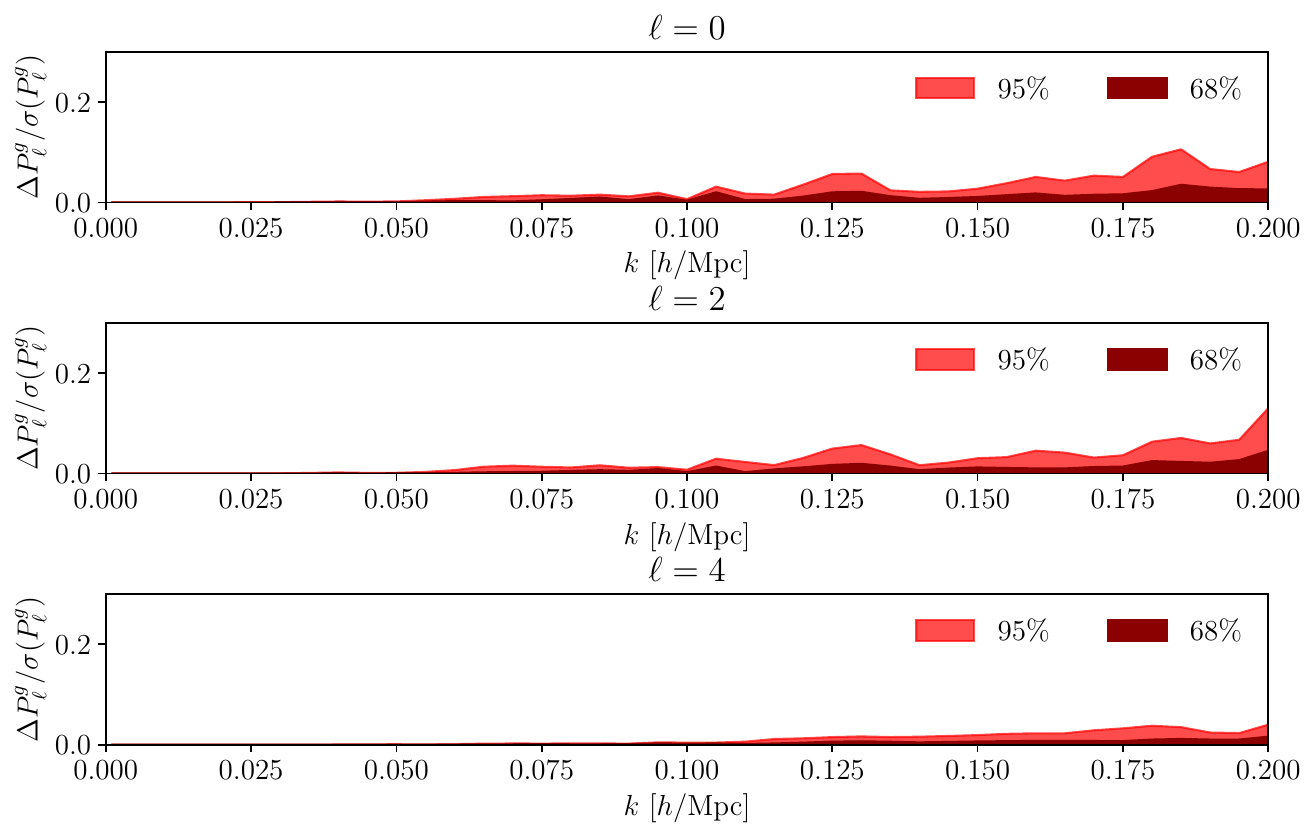}
      \includegraphics[width=0.31\textwidth]{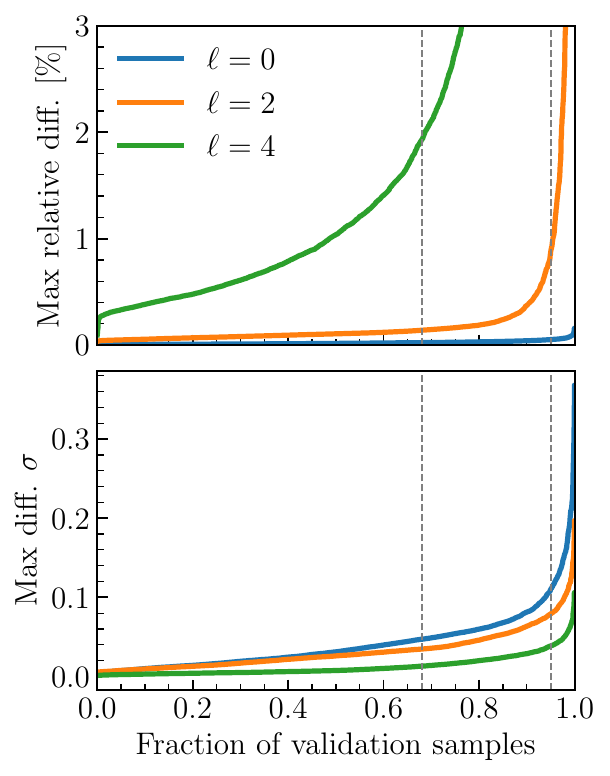}
    \caption{\textbf{Emulator accuracy} --- \textit{Left panel}: 68\% and 95\% quantiles of the differences in the galaxy power spectrum multipoles across scales, computed with \PyBird. 
    The comparison is between the NN-based emulator predictions and the full calculations, evaluated over the independent validation set described in sec.~\ref{sec:performance}. 
    Errors are shown relative to representative uncertainties expected for Stage-4 LSS surveys as described in the text. 
    \textit{Right panel}: Cumulative histogram of the maximum absolute differences across the full range of scales, with the two vertical dashed lines indicating the 68\% and 95\% quantiles of the validation samples.
    }
    \label{fig:emulator_residue}
\end{figure} 

\paragraph{Accuracy}
While full end-to-end precision tests of $\JAXEmu$ on recovered cosmological parameters are presented in sec.~\ref{sec:applications}, we first assess the emulator's accuracy using an independent validation set constructed as follows. 
Note that this validation set is different than the \texttt{CosmoRef} bank that we used to inform the training of our emulator. 
We adopt the same cosmological parameter space used to train \texttt{COMET}~\cite{Eggemeier:2022anw}, an emulator for the galaxy power spectrum in redshift space developed to meet Euclid’s precision requirements. 
These cosmologies are drawn from the following flat priors: $\omega_{\rm b} \in [0.02050 , 0.02415]$, $\omega_{\rm cdm} \in [0.085, 0.155]$, $n_s \in [0.92, 1.01]$, $\sigma_{12} \in [0.2, 1.0]$, and $f \in [0.5, 1.05]$, 
where $\sigma_{12}$ denotes the amplitude of fluctuations in spheres of radius $12$ Mpc. 
Crucially, evolution parameters such as $A_s$, $w_0$, $w_a$, and $\Omega_k h^2$ (as opposed to shape parameters such as $\omega_{\rm b}, \omega_{\rm cdm}$, or $n_s$), as well as the redshift $z$ dependence, affect the linear power spectrum in a manner equivalent to a rescaling of $\sigma_{12}$~\cite{Sanchez:2020vvb}. This means that the resulting test bank covers wide range of cosmological evolution scenarios and therefore provides a stringent test of the emulator.
We generate $5 \times 10^3$ such linear matter power spectra, which are then passed through both \JAXBird and \JAXEmu for comparison. 

Figure~\ref{fig:emulator_residue} shows the distribution of fractional errors on the independent validation set, normalised by representative observational uncertainties for Stage-4 LSS surveys. 
The survey characteristics assumed are an effective volume of $V_{\rm eff} = 50\,\mathrm{Gpc}^3$ and a tracer number density of $\bar{n} = 5 \times 10^{-4}\,(\hinvMpc)^3$. 
These uncertainties are estimated under the Gaussian approximation, assuming a fiducial Kaiser power spectrum at the \textit{Planck} best fit cosmology with linear bias $b_1 \sim 2$, evaluated at the effective redshift $z_{\rm eff}$ corresponding to each tested spectrum. 
Note that $V_{\rm eff} = 50\,\mathrm{Gpc}^3$ is representative of the total effective volume of a Stage-4 LSS survey, while the volume per redshift bins in practice is usually less than $10 \,\mathrm{Gpc}^3$ (see table~[2] of our companion paper for DESI-like survey characteristics per redshift bins). 
The emulators meet the accuracy requirements comfortably across the full parameter space, with 95\% of the residuals remaining below $0.1\sigma$ for all multipoles. 
Here we have focused on assessing emulator accuracy within the regime where the one-loop theoretical model for the redshift-space galaxy power spectrum is expected to be reliable, namely up to $k_{\rm max} = 0.2\,\hinvMpc$. 
As data precision increases, the theoretical $k$-reach is expected to decrease. 
Consequently, as survey volumes grow, the emulator is expected to remain relevant, as its accuracy improves at larger scales (lower $k$). 
Nevertheless, in appendix~\ref{app:emuacc}, we provide additional results on emulator accuracy across a broader range of scales, extending to $k_{\rm max} = 0.3\,\hinvMpc$. 
There, we observe that the relative error increases, reaching approximately $0.5\sigma$. 
However, we do expect that the induced bias on cosmological parameters to be smaller, as (\textit{i}) the induced error on the total $\chi^2$ is a mean square error averaged over all $k$'s, and (\textit{i}) numerical errors tend to average out in posterior sampling. 
Indeed, as shown in sec.~\ref{sec:PT}, when applied to large-volume simulations, the resulting bias on inferred cosmological parameters, when fitting the data in wedges up to $k_{\rm max} = 0.3\,\hinvMpc$, remains tolerably small in practice.

\paragraph{Speed}
Table~\ref{tab:speed} compares the wall-clock time required to evaluate the three redshift-space multipoles of the galaxy power spectrum with different back-ends and hardware configurations.  
Switching from \texttt{PyBird} to its \texttt{JAX} rewrite already provides a $5\times$ speed-up on a CPU through just-in-time (JIT) compilation. Replacing the costly loop integrals and IR-resummation with NN-based emulators brings a gain of two order-of-magnitude on the CPU, with another order moving to a Tesla A100 GPU. Furthermore, using \texttt{JAX}'s native \texttt{vmap} functionality, the vectorisation of the code enables another factor of $\sim 3-10$ speed up on the GPU hardware we used for benchmarking, depending on the input batch size (tested from 48 to 128). 

These runtime gains are broadly consistent with back-of-the-envelope counts of floating-point operations (flop).  For the full \texttt{PyBird} calculation, the dominant cost stems from the 22-diagram loop integrals, which is computed using the FFTLog as a matrix multiplication~\cite{Simonovic:2017mhp,DAmico:2020kxu,Zhang:2021yna}
\begin{equation}\label{eq:22}
P_{22}(k) = \sum_{n=1}^{N_{\rm FFT}}\sum_{m=1}^{N_{\rm FFT}} c_n k^{-2\nu_n} \cdot M_{22}(m,n) \cdot k^{-2\nu_m} c_m \ , 
\end{equation}
where $\nu_n$ are complex powers resulting from the FFTLog decomposition of $P_{\rm lin}$, and where $M_{22}(m,n)$ are matrices that are pre-computed.\footnote{Note that $P_{22}$ here is not $P^{ij}$ in eq.~\eqref{eq:master} with $i=2=j$. 
Instead, many $P^{ij}$ are of the 22-diagram type, and are computed following eq.~\eqref{eq:22} for their respective $M_{22}$.  } 
The correlation function, used in the IR-resummation scheme of \PyBird as detailed in app.~\ref{app:resum}, is computed likewise~\cite{DAmico:2020kxu,Zhang:2021yna}. 
The complexity of these operations can be approximated by
\[
  \mathcal{C}_{\text{loops}}
  \;\simeq\;
  \, N_{\rm FFT}^{2}\,N_k\,N_{\rm loop}
  \;.
\]  
With the default values FFTLog points \(N_{\rm FFT}=512\), length of the internal $k$-array \(N_k=77\), and number of 22-loop terms \(N_{\rm loop}=35\), this gives
\(
\mathcal{C}_{\text{loops}}\sim1\times10^{9}\;\text{flop}\,. 
\)
In practice, optimised matrix contraction through \texttt{NumPy einsum} leveraging \texttt{BLAS} backend reduces the flop counts by roughly an order-one factor. 
By contrast, a forward pass through the neural emulator involves
\[
  \mathcal{C}_{\text{emu}}
  \;\simeq\;
  N_{\rm multipoles}\bigl[N_{\rm in}N_{\rm nodes}
          +N_{\rm layers}N_{\rm nodes}^{2}
          +N_{\rm out}N_{\rm nodes}\bigr]
  \;,
\]
where \(N_{\rm multipoles}=3\) is the number of multipoles, \(N_{\rm in}=82\) is the number of input parameters, \(N_{\rm layers}=3\) the number of hidden layers, \(N_{\rm nodes}=512\) the width of each layer, and \(N_{\rm out}=256\) the number of retained PCA modes.  
Evaluating this expression yields
\(
\mathcal{C}_{\text{emu}}\sim2\times10^{6}\;\text{flop}\,.
\)
The \(\sim500\)-fold reduction in arithmetic work largely accounts for the observed timing improvement; the remaining discrepancy arises from \texttt{Python} overheads, memory traffic, and specific architecture designs, which are not captured by a simple flop estimate. 

\begin{table}[!ht]
  \centering
  \begin{tabular}{@{}l l c@{}}
    \toprule
    Architecture & \PyBird variant & Time per $3\times P^{\rm(\ell)}_{\,\!k}$ \\ \midrule
    CPU  & \PyBird\footnotemark    & 560\,ms \\
    CPU  & \JAXBird    & 102\,ms  \\
    CPU  & \JAXEmu     & 1.2\,ms   \\
    GPU  & \JAXBird    & 2.0\,ms   \\
    GPU  & \JAXEmu     & 0.19\,ms  \\
    GPU  & \JAXEmu (vectorised, batch = 128) & 0.023\,ms \\ \bottomrule
  \end{tabular}
  \caption{Wall-clock time to evaluate the monopole, quadrupole, and hexadecapole (\(3\times P^{(\ell)}_k\)) of the redshift-space galaxy power spectrum for different back-ends.}
  \label{tab:speed}
\end{table}
\footnotetext{Note that previous public version of \PyBird was using by default $N_{\rm FFT} = 256$, for which the corresponding timing is 285ms. See also app.~\ref{app:resum}. 
} 

\paragraph{Observational effects}
The evaluation time reported in tab.~\ref{tab:speed} are benchmarked for the IR-resummed one-loop power spectrum multipoles, without accounting for additional observational effects that need to be modelled to accurately fit the data. 
Following ref.~\cite{Bonici:2025ltp}, we upgrade the numerical scheme in \PyBird for correcting geometrical distortions from the Alcock-Paczynski (AP) effect to a Gauss-Lobatto-Legendre integration rule using 8 fixed points. 
This is enough to achieve a $10^{-5}$ relative precision compared to a reference quadrature integration, while making the runtime spent on correct the AP effects virtually negligible. 
As for the window function convolution (that in the case of periodic boxes reduces to a convolution with a binning matrix), it takes about $X-Y\%$ of the total runtime, depending on \PyBird back-ends.

\section{\JAXBird: boosted cosmological inference in LSS} \label{sec:interface}

\JAXBird is a differentiable code written in \Python-\JAX delivering fast evaluation of EFTofLSS predictions for galaxy-clustering observables and cosmological likelihoods of LSS probes. 
Providing as input a linear matter power spectrum and growth functions computable with a Boltzmann solver, \JAXBird outputs the nonlinear `one-loop' correlation functions of galaxies in redshift space, dressed with modeling of observational effects from a given survey. 
As a standalone tool, it can be used in combination with any Boltzmann code and samplers. 
Yet, we provide: 
\begin{itemize}
\item[-] An end-to-end pipeline to analyse $N$-point statistics of spectroscopic data: BOSS DR12 luminous red galaxies~\cite{BOSS:2015ewx,BOSS:2016wmc}, eBOSS DR16 quasars~\cite{eBOSS:2020mzp,eBOSS:2020yzd}, and simulated data of upcoming surveys. We intend to continuously add support for likelihoods of future data as they become available. 
\item[-] Built-in compatibility with existing Boltzmann solvers or emulators: \texttt{CLASS}\footnote{\url{http://class-code.net/}}~\cite{1104.2933}, \texttt{CosmoPower-JAX}\footnote{\url{https://github.com/dpiras/cosmopower-jax}}~\cite{Piras:2023aub}, \texttt{DISCO-EB}\footnote{\url{https://github.com/ohahn/DISCO-EB}}~\cite{Hahn:2023nvb}, and \texttt{Symbolic-Pk}\footnote{\url{https://github.com/DeaglanBartlett/symbolic_pofk}}~\cite{2311.15865,2410.14623}. 
\item[-] Built-in compatibility with several contemporary numerical samplers for cosmological inference: ensemble sampling from \texttt{emcee}\footnote{\url{https://emcee.readthedocs.io/}}~\cite{1202.3665} or \texttt{zeus}\footnote{\url{https://zeus-mcmc.readthedocs.io/}}~\cite{2105.03468}, Hamiltonian Monte Carlo (HMC) (see ref.~\cite{Betancourt:2017ebh} for a review) based on the No-U-Turn algorithm (NUTS)~\cite{Hoffman:2011ukg} from \texttt{BlackJAX}\footnote{\url{https://blackjax-devs.github.io/blackjax/}}~\cite{blackjax}, Microcanonical Hamiltonian Monte Carlo (MC-HMC)~\cite{Robnik:2022bzs,Robnik:2022bzs}  also from \texttt{BlackJAX}, and nested sampling from \texttt{Nautilus}\footnote{\url{https://github.com/johannesulf/nautilus}}~\cite{2306.16923}. 
\item[-] An embedding within the widely-used cosmological inference software \texttt{MontePython}\footnote{\url{https://github.com/brinckmann/montepython_public}}~\cite{Audren:2012wb,1804.07261}. 
We also plan to upgrade the current \PyBird integration~\cite{Lai:2024bpl} to \JAXBird within \texttt{desilike}\footnote{\url{https://desilike.readthedocs.io/}}, the official analysis pipeline of DESI, and to integrate \JAXBird into \texttt{cloe}\footnote{\url{https://github.com/cloe-org}}, the Euclid analysis pipeline (currently under development).
\end{itemize}  

\paragraph{}
At its core, \JAXBird is essentially the same code as \PyBird\footnote{\url{https://github.com/pierrexyz/pybird}} --- the \Python code for Biased tracers in redshift space~\cite{DAmico:2020kxu}, but updated to \JAX, with further acceleration from the NN-based emulator presented in sec.~\ref{sec:emulation}. 
For this first release, \JAXBird has been gauged and tested for the one-loop power spectrum of galaxies in redshift space~\cite{Perko:2016puo}. 
We intend to extend \JAXBird to other cosmological observables, in particular the ones currently supported by \PyBird: one-loop bispectrum~\cite{DAmico:2022osl,DAmico:2022ukl} and two-point correlation function in configuration space~\cite{Zhang:2021yna}. 

\paragraph{}
A word on the implementation. Due to \JAX being literally composable transformations of a native \Python and \texttt{NumPy}\footnote{\url{https://numpy.org/}} program into a differentiable, vectorised, and GPU-compatible compiled version, \JAXBird is turned on with a simple backend switch on top of \PyBird, with the main code being agnostic on whether functions are called from the \JAX ecosystem or standard \Python library.
In particular, functions that differ in their call between their \Python and \JAX version are wrapped within single instances called by the main code of \PyBird, such that the latter is unique for both versions. 
Users and developers of \PyBird will enjoy all features brought by \JAX, that we detail in the following, with no particular adjustment other than installing extra dependencies. 
As a first introduction to \PyBird and its accelerating variants, we provide a general demonstration notebook here \href{https://github.com/pierrexyz/pybird/blob/master/demo/correlator.ipynb}{\faGithub}. 

\subsection{\JAX-powered features}\label{sec:jax}

\paragraph{Just-In-Time compilation}
A key strength of \JAX lies in its ability to remain fully compatible with the broader \Python environment while enabling Just-In-Time (JIT) compilation.\footnote{\url{https://docs.jax.dev/en/latest/jit-compilation.html}}
In typical cosmological inference workflows where \JAXBird is called repeatedly (see examples in sec.~\ref{sec:applications}), both \JAXBird and target likelihoods can be JIT compiled once at the beginning. 
This ensures that subsequent core computations are executed directly at the machine level, bypassing overheads of interpreting higher-level codes.
JIT compilation brings substantial speedups, as demonstrated in sec.~\ref{sec:performance}, particularly when run on a GPU.
A typical JIT compilation of \JAXBird completes in a few seconds, with additional time depending on the specific Boltzmann solver used in conjunction. 

\paragraph{Neural network embedding}
Although the neural network weights are trained using \texttt{TensorFlow} (as described in sec.~\ref{sec:emulation}), we embed the trained model smoothly into the \JAX ecosystem using the \texttt{flax}\footnote{\url{https://flax.readthedocs.io/}} library.
This integration allows our emulator-enhanced \JAXBird to be fully JIT-compiled, with automatic differentiation operating swiftly through the NN architecture.
The speed gains enabled by the emulator were presented in sec.~\ref{sec:performance}, while its accuracy in cosmological inference is evaluated in sec.~\ref{sec:applications}.

\paragraph{Boltzmann solver streaming}
While an increasing number of Boltzmann solvers or emulators are becoming \JAX-compatible ---enabling seamless integration with \JAXBird as we provide --- benchmark codes such as \texttt{CLASS}~\cite{1104.2933} and \texttt{CAMB}\footnote{\url{https://camb.info/}}~\cite{Lewis:1999bs,Howlett:2012mh} are not.
These established solvers continue to offer unmatched precision and a comprehensive set of cosmological models for exploration.
However, being written in \texttt{C} and interfaced with \Python via \texttt{Cython}\footnote{\url{https://cython.org/}}, \texttt{CLASS} cannot be directly embedded into the \JAX ecosystem\footnote{This may be possible in the future with tools such as \texttt{Enzyme-JAX}: \url{https://github.com/EnzymeAD/Enzyme-JAX}}.
To retain a fully JIT-compiled and differentiable pipeline when using such solvers, \JAXBird can switch to a finite-difference-based Taylor expansion of the observables, as detailed in sec.~\ref{sec:AD}. 
Built-in compatibility between \JAXBird and Boltzmann solvers are showcased here \href{https://github.com/pierrexyz/pybird/blob/master/demo/correlator.ipynb}{\faGithub}. 

\paragraph{Vectorised ensemble sampling}
\JAXBird, when combined with a Boltzmann solver, computes predictions for LSS likelihoods based on input cosmological parameter sets.
Thanks to \JAX automatic vectorisation via \texttt{vmap}\footnote{\url{https://docs.jax.dev/en/latest/automatic-vectorization.html}}, \JAXBird can efficiently process batches of input parameters. 
This leads to significant speedups over sequential evaluation by leveraging optimised matrix algebra from low-level libraries such as \texttt{BLAS}, while also minimising overheads.
Vectorisation is particularly advantageous for ensemble sampling, allowing multiple walkers to run in parallel.
For \JAXBird, we observe favorable time scaling with batch sizes up to $\sim 1024$. 
In practice, using around $\sim 128$ walkers offers a good trade-off: rapid convergence to the target distribution is achieved with a manageable number of burn-in steps and efficient sampling is obtained thereafter in a few steps.
An example of cosmological inference with \JAXBird using vectorised ensemble sampling is provided here \href{https://github.com/pierrexyz/pybird/blob/master/demo/inference.ipynb}{\faGithub}. 

\subsection{AD-powered features}\label{sec:AD}
\JAX enables machine-precision differentiation with modest overhead via automatic differentiation (AD). 
AD exploits the fact that a program is a composition of elementary operations with known derivatives. 
By recording the computation graph (``\emph{evolution tracing}'') and applying the chain rule, one obtains derivatives of the full composition (see ref.~\cite{1502.05767}). For scalar objectives of many parameters (\textit{e.g.}, likelihoods), reverse-mode AD (backpropagation) is typically optimal, propagating sensitivities from outputs back to inputs.\footnote{On modern hardware such as GPUs, the simpler primitives of forward-mode AD can sometimes compensate for the extra evaluations; see, \textit{e.g.}, ref.~\cite{2405.12965}.} Reverse-mode proceeds in two passes: a forward pass that builds the trace, and a backward pass that accumulates gradients. In brief, AD in \JAXBird provides fast, accurate derivatives of EFTofLSS predictions and their likelihoods. The new AD-enabled functionalities are summarised below.

\paragraph{Fisher matrix}
With AD, estimating Fisher matrix as the Hessian of the log-likelihood functions of target experiments has never been easier --- a one-liner code for a numerically stable computation. 
Beyond the common use of forecasting parameter constraints, the Fisher matrix can be leveraged to increase efficiency of numerical sampling by starting at positions drawn from a multivariate Gaussian centered on some fiducials (\textit{e.g.}, the maximum a posteriori) with covariance given by the inverse-Fisher matrix, thereby reducing the burn-in phase. 
For a typical $\mathcal{O}(10)$-parameter likelihood in galaxy clustering, \JAXBird computes the Fisher matrix in seconds (with the timing largely dominated by JIT compilation), an amount of time worth spending to initialise subsequent sampling, reducing significantly the burn-in time. 
An example is provided here \href{https://github.com/pierrexyz/pybird/blob/master/demo/run.ipynb}{\faGithub}. 
The Fisher matrix also enters the non-flat integration measure used to define unbiased expectation values as shown in our companion paper~\cite{paper2} (also detailed below). 

\paragraph{Taylor expansion of observables}
Observables in LSS such as the power spectrum and bispectrum are smooth functions of cosmological parameters. 
Their cosmological dependence can therefore be Taylor-expanded around a fiducial cosmology, enabling extremely fast emulation of theoretical predictions~\cite{Cataneo:2016suz,Colas:2019ret}. 
With AD, \JAXBird can precompute in minutes the required Taylor expansions for the observables entering the target likelihoods, allowing for highly accelerated sampling. 
When the Boltzmann solver is not itself differentiable, \JAXBird switches to finite difference, and the resulting emulator from the Taylor expansion is then effectively differentiable. 
Accuracy benchmarks are reported in sec.~\ref{sec:sampler}, and the corresponding implementation is showcased here 
\href{https://github.com/pierrexyz/pybird/blob/master/demo/inference.ipynb}{\faGithub}.

\paragraph{Gradient-based optimisation}
The minimum $\chi^2$ is the standard metric used to assess the quality of a model's fit to experimental data. 
Correspondingly, the maximum a posteriori (MAP) estimate, or posterior mode, is commonly reported as the most likely point estimate for model parameters under a given likelihood and prior. 
Finding the MAP boils down to solving an optimisation problem over a scalar objective function, which is typically non-analytic and must be approached numerically.
In high-dimensional or multimodal parameter spaces, gradient information becomes essential for efficient exploration. 
This is particularly relevant for galaxy-clustering likelihoods, where the number of nuisance parameters scales with the number of observed sky patches --- easily reaching $\mathcal{O}(100)$ parameters in current Stage-4 LSS surveys. 
Additionally, increasingly sophisticated modeling compounds this complexity.
With \JAXBird, we find that the classic \texttt{MIGRAD} algorithm, as implemented in the widely used \texttt{Minuit}\footnote{\url{https://scikit-hep.org/iminuit/}} minimiser~\cite{James:1975dr}, performs reliably in low-dimensional settings ($\lesssim 10$ parameters), especially for near-Gaussian posteriors as typically encountered in cosmology. 
However, it often fails to locate the MAP as the number of parameters increases --- either due to added sky patches or more complex models --- particularly when poorly constrained directions lead to nearly flat regions in parameter space, \textit{i.e.}, large parameter degeneracies.
In contrast, \texttt{ADAM}\footnote{as implemented in \url{https://optax.readthedocs.io/}}~\cite{1412.6980}, a first-order gradient-based stochastic optimiser, consistently performs well across all practical cases tested with \JAXBird (see sec.~\ref{sec:applications}).
We anticipate that its performance can surpass that of global optimisers such as simulated annealing, as implemented in \texttt{Procoli}~\cite{2401.14225} and \texttt{PROSPECT}~\cite{2312.02972}, both in terms of robustness and convergence speed. 
We leave a detailed comparison for future work. 
Typically, \texttt{ADAM} converges within minutes for models with $\mathcal{O}(10)$ parameters in galaxy-clustering applications using \JAXBird. 
Because point estimates like the MAP can be sensitive to numerical noise, we recommend computing them without using the emulator, which introduces small but non-negligible errors at each point in parameter space. These minor inaccuracies can shift the MAP estimate, especially in flat regions of the posterior. 
In contrast, over many evaluations, such stochastic errors tend to average out, making the emulator well suited for sampling-based inference and accurate estimation of full posterior distributions, that we now detail. 

\paragraph{Gradient-based sampling} 
To explore increasingly large parameter spaces—such as those encountered in galaxy clustering analyses that incorporate multiple tracers across different sky patches and redshift bins, each contributing their own set of nuisance parameters—gradient-based sampling offers a practical alternative to traditional samplers, which often do not scale efficiently with dimensionality.
Performance of \texttt{HMC-NUTS} for a $\mathcal{O}(10$–$100)$-parameter case are compared against ensemble sampling in sec.~\ref{sec:applications}. 
In practice, the total runtime depends on several factors, including the tuning of hyperparameters, the length of the burn-in period, and critically, the additional cost of computing gradients, which we find to be around $3\times$ the likelihood call. 
This makes gradient-based sampling a method of choice, as demonstrated in sec.~\ref{sec:sampler} and here \href{https://github.com/pierrexyz/pybird/blob/master/demo/run.ipynb}{\faGithub}.

\paragraph{Marginals with non-flat volume measure}
Perhaps among the additional features enabled by AD, this one stands out for its novelty.
As demonstrated in our companion paper~\cite{paper2}, we propose to improve parameter estimation by redefining our expectation values with respect to non-flat volume measures, removing bias in marginal posteriors known as volume projection effects (see \textit{e.g.}, refs.~\cite{Ivanov:2019pdj,DAmico:2022osl,Carrilho:2022mon,Simon:2022lde,DESI:2024jis,Paradiso:2024yqh}). 
For a smooth posterior distribution $\mathcal{P}(\pmb{\theta}|y)$ over model parameters $\pmb{\theta}$ obtained from the data $y$ that admits a global maximum, the optimal volume measure we put forward takes a form close to the well-known Jeffreys \emph{prior}~\cite{paper2}, 
\begin{equation}\label{eq:measure}
\mathcal{M}_{\mathcal{H}}(\pmb\theta) = \sqrt{\det \mathcal{H}(\pmb \theta)} \, d^N\pmb \theta \, , \qquad \mathcal{H}_{\mu\nu}(\pmb \theta) = \partial_\mu\partial_\nu \log \mathcal{P}(\pmb \theta|y=m(\pmb {\theta_*})) \, .
\end{equation}
Here $\mathcal{H}_{\mu\nu}$ is a matrix taken as the Hessian over the distribution $\log \mathcal{P}(\pmb \theta|y=m(\pmb {\theta_*}))$, where the data are taken to be the model $m$ evaluated on the mode $\pmb{\theta_*}$. 
In practice, we can add to each sample of the log-posterior obtained under the standard flat measure a log-measure weight to compute marginal statistics. 
This measure is found to correct for the average mean bias we derive at leading order in the Laplace expansion of the posterior~\cite{paper2}, 
\begin{equation}\label{eq:bias}
b [\mathcal{F}, \pmb {\theta_*}] = (\mathcal{F}^{-1}_{\alpha\mu} \mathcal{F}^{-1}_{\nu\rho} \partial_\rho \mathcal{F}_{\mu\nu})\big|_{\pmb \theta = \pmb{\theta_*}}
\end{equation}
Alternatively, this bias can be computed once and used as post-debiasing correction to shift the posterior, yielding an unbiased representation of marginal distributions and their credible intervals. 
With the upgrade to \JAXBird, both the log-measure weight~\eqref{eq:measure} or the bias term~\eqref{eq:bias} can now be computed seamlessly and at machine precision, making the construction of unbiased marginal inferences in large-scale structure efficient and robust. 
Since this paper focuses on the speed and accuracy performance of \JAXBird, we do not apply this correction in the applications discussed in sec.~\ref{sec:applications}.
Practical examples of its use are instead provided in our companion paper~\cite{paper2} with implementation showcased here \href{https://github.com/pierrexyz/pybird/blob/master/demo/inference.ipynb}{\faGithub}.

\section{Cosmological applications}\label{sec:applications}

To demonstrate the accuracy and flexibility of the cosmology-independent emulator of \JAXBird presented in sec.~\ref{sec:emulation}, that we dub \JAXEmu, we compare posteriors of cosmological parameters from LSS analysis obtained with or without the use of the emulator. 
In sec.~\ref{sec:PT}, we fit large-volume simulations to assess the accuracy of \JAXEmu in recovering cosmological parameters. 
In sec.~\ref{sec:BOSS}, we fit the Sloan Digital Sky Survey Baryon Oscillation Spectroscopic Survey (SDSS BOSS)~\cite{BOSS:2016wmc} Luminous Red Galaxies (LRGs)~\cite{BOSS:2015ewx} of modest volume compared to ongoing and upcoming surveys such as 
DESI~\cite{DESI:2024aax} or Euclid~\cite{Euclid:2024yrr}, making it ideal to stress test \JAXEmu's coverage in parameter and model space.  
Finally in sec.~\ref{sec:sampler}, we perform a forecast on a DESI-like survey consisting of 7 individual skies, each fit with its unique EFT nuisance parameters to capture variations in the selection function and redshift, making a total of $84$ parameters to explore on top of the cosmological parameters of interest. 
This provides an ideal realistic setup to test the various sampling strategies presented in sec.~\ref{sec:interface}. 
Taken together, this suite of analyses aims to demonstrate the raw precision, coverage in parameter and model space, and efficiency of \JAXEmu in inferring cosmological parameters from the incoming data of Stage-4 spectroscopic surveys in the next decade. 

\begin{table}[htbp]
    \centering
    \begin{tabular}{l@{\hspace{2cm}}l@{\hspace{2cm}}l}
        \hline
        Category & Parameter & Prior \\
        \hline
        \multirow{4}{*}{Galaxy biases} 
            & $b_{1}$ & $[0, 5]$ \\
            & $b_{2}$ & $\mathcal{N}(0, 5)$ \\
            & $b_{3}$ & $\mathcal{N}(0, 5)$ \\
            & $b_{4}$ & $\mathcal{N}(0, 5)$ \\
        \hline
        \multirow{3}{*}{Counterterms} 
            & $c_{\rm ct}$ & $\mathcal{N}(0, 5)$ \\
            & $c_{r,1}$ & $\mathcal{N}(0, 5)$ \\
            & $c_{r,2}$ & $\mathcal{N}(0, 5)$ \\
        \hline
        \multirow{3}{*}{Stochastic terms} 
            & $c_{\epsilon,0}$ & $\mathcal{N}(0, 2)$ \\
            & $c_{\epsilon,1}$ & $\mathcal{N}(0, 5)$ \\
            & $c_{\epsilon,2}$ & $\mathcal{N}(0, 5)$ \\
        \hline
        \multirow{2}{*}{NNLO counterterms} 
            & $c_{r,4}$ & $\mathcal{N}(0, 5)$ \\
            & $c_{r,6}$ & $\mathcal{N}(0, 5)$ \\
        \hline
    \end{tabular}
    \caption{\textbf{Priors for EFT nuisance parameters used in all analyses} ---  
    All are assigned Gaussian priors $\mathcal{N}(\mu, \sigma)$, with $\mu$ denoting the central value and $\sigma$ the standard deviation, except for $b_1$, which is sampled with a flat prior. 
    These priors, inspired by refs.~\cite{DAmico:2019fhj,Simon:2022lde,DESI:2024jis}, are based on order-of-magnitude estimates for the parameter sizes, motivated by naturalness considerations to ensure that EFTofLSS predictions remain within the perturbative regime where the theory is valid. 
    For the EFT scales associated with the counterterms, we adopt $k_{\rm NL} = k_{\rm M} = 0.7\,h\,\mathrm{Mpc}^{-1}$ and $k_{\rm R} = 0.25\,h\,\mathrm{Mpc}^{-1}$. 
    See main text for a detailed summary of the EFT parameters and their role in modeling the galaxy power spectrum in redshift space at one loop in the EFTofLSS.
    }
    \label{tab:priors}
\end{table}

\paragraph{Cosmological inference setup} 
For all analyses of galaxy-clustering data presented in this section, as well as in our companion paper~\cite{paper2}, we work with the following inference setup. 
We fit a data vector $y$ consisting of the first three even power spectrum multipoles ($\ell = 0,2,4$) concatenated together in $k$-bins falling in the range $[k_{\rm min}, k_{\rm max}]=[0.01, 0.20] \, \hinvMpc$, except if stated otherwise. 
We use a Gaussian likelihood to describe the data $y$, 
\begin{equation}\label{eq:likelihood}
-2 \log \mathcal{L}(y | \pmb{\theta}) = (m(\pmb{\theta}) - y)^T C^{-1} (m( \pmb{\theta}) - y) \ ,
\end{equation}
where $m(\pmb \theta)$ is the model (summarised below) evaluated with \PyBird or its variants considered in this work, with $\pmb \theta$ the model parameters consisting in the cosmological parameters of interest and the EFT parameters describing each sky composing the dataset considered. 
Here $C^{-1}$ is the inverse covariance whose estimation is described below for each dataset considered in our works. 
To estimate the posterior of $\pmb \theta$, we run MCMC chains using various samplers described in sec.~\ref{sec:interface}. 
When not stated otherwise, we use the Metropolis-Hasting algorithm implemented in \texttt{MontePython}~\cite{Audren:2012wb,1804.07261}, running $8-16$ chains in parallel, with convergence monitored using the Gelman-Rubin criterion $R < 0.01$. 
All triangle plots in this paper are produced using \texttt{GetDist}\footnote{\url{https://getdist.readthedocs.io}}~\cite{1910.13970}. 

\paragraph{Model and priors}
The one-loop model for redshift-space galaxy power spectrum that we consider in this work depends on twelve EFT parameters: 
\begin{itemize}
    \item \textit{Galaxy biases}: four parameters $b_i, i=1,\dots, 4$, entering the perturbation theory loop kernels of galaxies in redshift space, as formulated in the Basis of Descendants~\cite{Angulo:2015eqa,Fujita:2016dne} (see also refs.~\cite{McDonald:2009dh,Assassi:2014fva,Mirbabayi:2014zca,Eggemeier:2018qae,DAmico:2022ukl}). 
    \item \textit{Counterterms}: three parameters $c_{\rm ct}, c_{r,1}, c_{r,2}$, renormalising the UV-divergence of the $P_{13}$ loop diagram~\cite{Carrasco:2012cv,Pajer:2013jj,Senatore:2014vja,Perko:2016puo}. These scale relative to $P_{\rm lin}$ as $\frac{k^2}{k_{\rm M}^2}$, and $\frac{k^2}{k_{\rm R}^2} \mu^2, \frac{k^2}{k_{\rm R}^2} \mu^4$, with $k_{\rm M}^{-1}$ and $k_{\rm R}^{-1}$ denoting characteristic EFTofLSS scales controlling the spatial derivatives or redshift-space velocity products~\cite{Senatore:2014eva,DAmico:2021ymi}.
    \item \textit{Stochastic terms}: three parameters $c_{\epsilon,0}, c_{\epsilon,1}, c_{\epsilon, 2}$, renormalising the UV-divergence of the $P_{22}$ loop diagram and scaling (in units of the mean number density $\bar n^{-1}$) as $k^0$, $\frac{k^2}{k_{\rm M}^2}$, and $\frac{k^2}{k_{\rm M}^2}\mu^2$, respectively~\cite{Perko:2016puo}. 
    \item \textit{Next-to-next-leading order redshift-space counterterms}: two parameters $c_{r,4}, c_{r,6}$, enhanced given a large $k_{\rm R}^{-1}$ (compared to $k_{\rm M}^{-1} \sim k_{\rm NL}^{-1}$), scaling relatively to $P_{\rm lin}$ as $\frac{k^4}{k_{\rm R}^4}\mu^4$ and $\frac{k^4}{k_{\rm R}^4}\mu^6$~\cite{DAmico:2021ymi}. 
\end{itemize}
Our priors on the EFT parameters consist in large Gaussian distributions specified in table~\ref{tab:priors}. 
All EFT parameters—except the galaxy biases ${b_1,b_2,b_3}$—enter the model linearly and are analytically marginalised as detailed in App.D of our companion paper\cite{paper2}.
Cosmological parameters are scanned with large flat priors if not stated otherwise. 
We sample the $\Lambda$CDM parameters $\lbrace \omega_{\rm b}, \omega_{\rm cdm}, \ln(10^{10}A_s)$, $n_s\rbrace$, with eventually extra parameters in the model extensions presented below. 
We however choose to present posteriors for the fractional total matter abundance $\Omega_m$ and the clustering amplitude $\sigma_8$ instead of $\omega_{\rm cdm}$ and $\ln(10^{10}A_s)$ to meet the conventions in the LSS community. 
When analysing the BOSS data, we consider one neutrino at minimal mass $m_\nu = 0.06 \,$eV and two massless following \textit{Planck} prescription~\cite{Planck:2018vyg} (see \textit{e.g.}, refs.~\cite{2003.03354,Racco:2024lbu} for the validity of this approximation). 
When analysing simulations of synthetic data, we do not vary neutrinos.

\subsection{PT challenge simulations}\label{sec:PT}

To gauge the accuracy of \JAXEmu, we make use of the PT challenge mocks, a suite of 10 high-resolution $N$-body simulations painted with a high-fidelity Halo Occupation Distribution (HOD) model, described in ref.~\cite{Nishimichi:2020tvu}. 
By fitting the average power spectrum obtained from combining measurements across individual boxes, corresponding to a total data volume of $566 \, (\textrm{Gpc}/h)^3$, the PT challenge offers a stringent accuracy test for theoretical prediction codes in galaxy clustering. 
This setup enables direct comparison to the (blinded) truth of the simulations, as originally designed for testing predictions from the EFTofLSS~\cite{Nishimichi:2020tvu}.\footnote{Submitted results are available at~\href{https://www2.yukawa.kyoto-u.ac.jp/~takahiro.nishimichi/data/PTchallenge/}{here}.} 
In the original challenge presented in ref.~\cite{Nishimichi:2020tvu}, a pre-version of \PyBird was used, showing that the power spectrum multipoles could be fit up to $k_{\rm max} \simeq 0.14 \, h/\textrm{Mpc}$ reliably for the given simulation volume without significant bias in the inferred cosmological parameters. 
In ref.~\cite{DAmico:2021ymi}, it was further demonstrated using \PyBird that higher $k_{\rm max}$ could be reached by constructing a linear combination of the power spectrum multipoles dubbed $\slashed{P}$ where the terms proportional to powers of $\mu$, arising from the redshift-space distortions, are killed (see also ref.~\cite{Ivanov:2021fbu}). 
In short, the redshift-space distortions are controlled by a large renormalisation scale $\sim 1/k_{\rm R}$ (see also ref.~\cite{Lewandowski:2015ziq}). 
In comparison, the EFTofLSS prediction for $\slashed{P}$ is roughly controlled by the nonlinear scale $1/k_{\rm NL}$ (and the galaxy spatial extension $1/k_{\rm M}$), which is typically a few times smaller than $1/k_{\rm R}$. 
Thanks to an increased convergence, the $k$-reach in the fit to $\slashed{P}$ can be pushed beyond the one of the traditional multipoles, typically up to $k_{\rm max} \simeq 0.30 \, h/\textrm{Mpc}$ on the PT challenge as demonstrated in ref.~\cite{DAmico:2021ymi}. 
The fit to $\slashed{P}$ combination from the PT challenge power spectrum thus makes an highly demanding accuracy test for \JAXEmu covering all modes up to the highest $k$s reachable in current galaxy clustering survey with the EFTofLSS at one loop. 

\begin{figure}[!ht]
  \centering
  \includegraphics[width=0.49\textwidth]{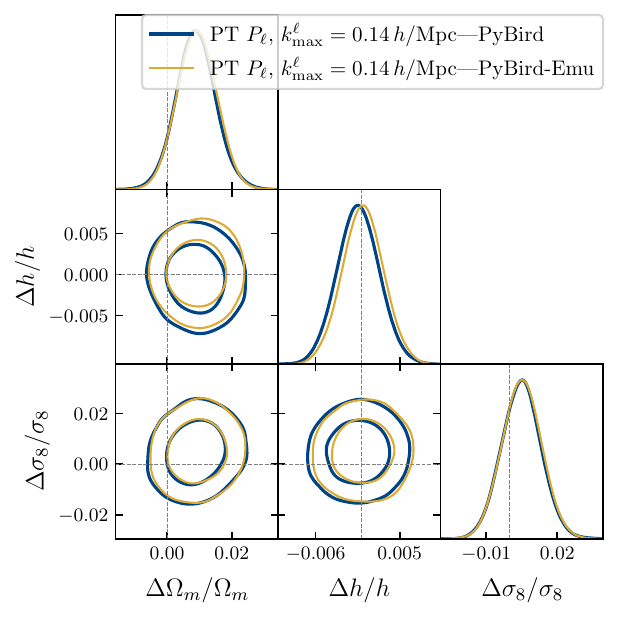}
  \includegraphics[width=0.49\textwidth]{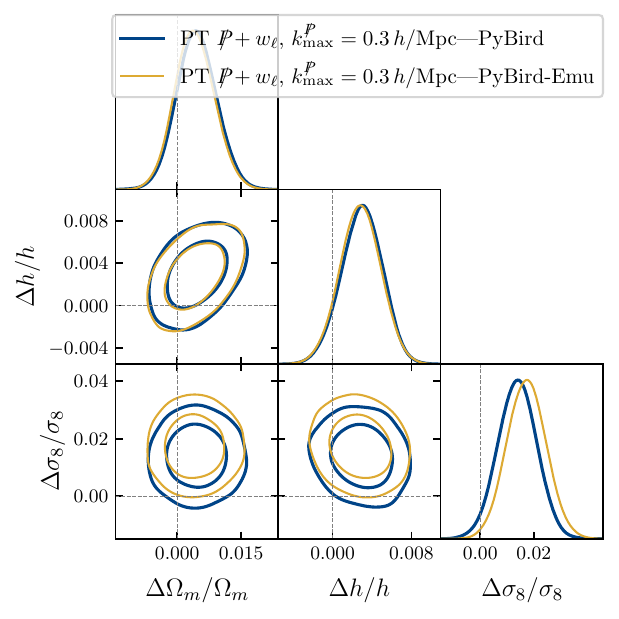}
  \caption{\textbf{Comparison \PyBird vs. \JAXEmu on PT simulation data} --- 1D and 2D marginal posterior distributions of inferred $\Lambda$CDM parameters from the PT challenge simulations, with fixed $\omega_{\rm b}$ and $n_s$. 
  For all configurations, all parameters are recovered within $\sim 1\sigma$, with the truth shown in dashed line. 
  The posteriors from \PyBird (\textit{blue contours}) and \JAXEmu (\textit{yellow contours}) agree at subpercent level ($<0.3\%$ of the parameter values on 1D marginals) whether fit in multipoles $P_\ell$ up to $k^\ell_{\rm max} = 0.14 \hinvMpc$ (\textit{left panel}) or in wedges $\slashed{P} + w_\ell$ with $\slashed{P}$ analysed up to $k^\slashed{P}_{\rm max} = 0.3 \hinvMpc$ (\textit{right panel}). 
  This validates the raw accuracy of \JAXEmu in recovering cosmological parameters to high precision. 
  }
  \label{fig:PT}
\end{figure}

In fig.~\ref{fig:PT}, we show cosmological results obtained by fitting either the power spectrum multipoles $P_\ell$ (using $k_{\rm max} = 0.14 \, h/\textrm{Mpc}$) or $\slashed{P}$ (using $k_{\rm max}^{\slashed{P}} = 0.30 \, h/\textrm{Mpc}$) and two wedges $w_1, w_2$ (using $k_{\rm max}^{1/2} = 0.20, 0.10 \, h/\textrm{Mpc}$; see ref.~\cite{DAmico:2021ymi} for precise definition of these combinations and justifications for the choice of $k_{\rm max}$s). 
We see that \JAXEmu perform equally well than \PyBird in recovering the true cosmological parameters within uncertainties, with their posteriors differing less than $0.06\%$ for all parameters in both cases, with the exception of $\sigma_8$ from $\slashed{P} + w_{1,2}$ that differs at $\sim 0.3\%$. 
These represent negligible errors with respect to target precision in cosmology in the next decade (of the order $\sim 1\%$). 
\JAXEmu is thus sufficiently accurate in recovering cosmological parameters from the full-shape power spectrum of ongoing and upcoming galaxy surveys.

\subsection{BOSS data}\label{sec:BOSS}
Having validated the raw accuracy of \JAXEmu, we now assess its performance in regards of two other metrics: coverage across cosmological parameter space and flexibility in reliably fitting extended cosmological models. 

\paragraph{Data and likelihoods}
Our main dataset is built from the BOSS DR12 LRG samples~\cite{BOSS:2015ewx}. 
The power spectrum multipoles, obtained in ref.~\cite{DAmico:2022osl}, were measured using the estimators developed in refs.~\cite{Feldman:1993ky,Yamamoto:2005dz,1312.4611,Bianchi:2015oia,Scoccimarro:2015bla,BOSS:2015npt} using the FFT-based code \texttt{Rustico}\footnote{\url{https://github.com/hectorgil/Rustico}}~\cite{BOSS:2015npt}, from the BOSS DR12 catalogs (v5) combined CMASS-LOWZ\footnote{Publicly available at \url{https://data.sdss.org/sas/dr12/boss/lss/}} described in ref.~\cite{BOSS:2015ewx}. 
The covariance is estimated as the scatter across measurements on the patchy mock suites~\cite{Kitaura:2015uqa}. 
We apply the Hartlap factor to correct bias from the finite number of mocks used to estimate the inverse covariance~\cite{Hartlap:2006kj}. 
We correct for the Alcock-Pazcynski effects and window functions as described in ref.~\cite{DAmico:2019fhj} (with the window, which formalism was developed in refs.~\cite{Wilson:2015lup,BOSS:2016psr,Beutler:2018vpe}, measured using \texttt{fkpwin}\footnote{\url{https://github.com/pierrexyz/fkpwin}}~\cite{Beutler:2018vpe}), and fiber collisions following ref.~\cite{Hahn:2016kiy}. 
The likelihood, priors, and posterior sampling are described above in the beginning of sec.~\ref{sec:applications}. 

We further make use of various external probes depending on the cosmological exploration. 
For all our BOSS analyses, we use a prior from Big-Bang Nucleosynthesis (BBN) experiments~\cite{Schoneberg:2019wmt,1712.04378,Cooke:2017cwo,Aver:2015iza}, that essentially constrain $\omega_{\rm b}$. 
We also sometimes use information from low-redshift supernovae from Pantheon+~\cite{2202.04077}, with magnitude either uncalibrated or calibrated with the S$H_0$ES distance ladder~\cite{2112.04510}.  

Due to its relatively modest volume (especially when compared to ongoing and upcoming Stage-4 surveys), the BOSS data serves as an ideal testbed for assessing \JAXEmu's coverage in both parameter and model space. 
Notably, BOSS has been widely used to constrain a broad range of cosmological models (see \textit{e.g.}, refs.~\cite{DAmico:2020kxu,Bottaro:2023wkd,Poudou:2025qcx,Calderon:2025xod,DAmico:2022gki,Piga:2022mge,Spaar:2023his,Taule:2024bot,Lu:2025gki,Glanville:2022xes,Lague:2021frh,Gonzalez:2020fdy,Allali:2021azp,Allali:2023zbi,Simon:2022ftd,Niedermann:2020qbw,Cruz:2023cxy,DAmico:2020ods,Smith:2020rxx,Simon:2022adh,Gsponer:2023wpm,Simon:2022csv}), including models that yield non-trivial departures from $\Lambda$CDM. 
Then, the question we can ask is, do the posterior distributions of cosmological parameters obtained with \JAXEmu and \PyBird remain consistent in extended scenarios where cosmologies, potentially significantly different from $\Lambda$CDM, are probed within the ranges allowed by BOSS?

\begin{figure}[!ht]
  \centering
  \subcaptionbox{$\Lambda$CDM\label{fig:lcdm}}
    [.49\linewidth]{\includegraphics[width=\linewidth]{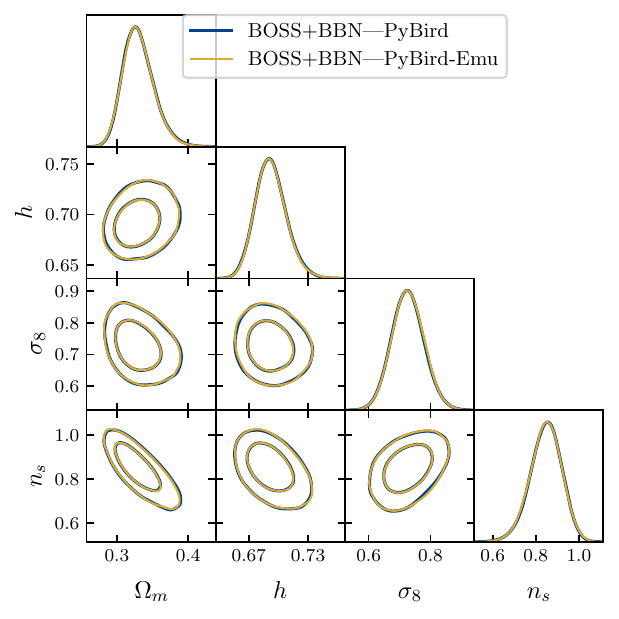}}\hfill
\subcaptionbox{$w_0w_a$CDM\label{fig:w0wa}}[.49\linewidth]{\includegraphics[width=\linewidth]{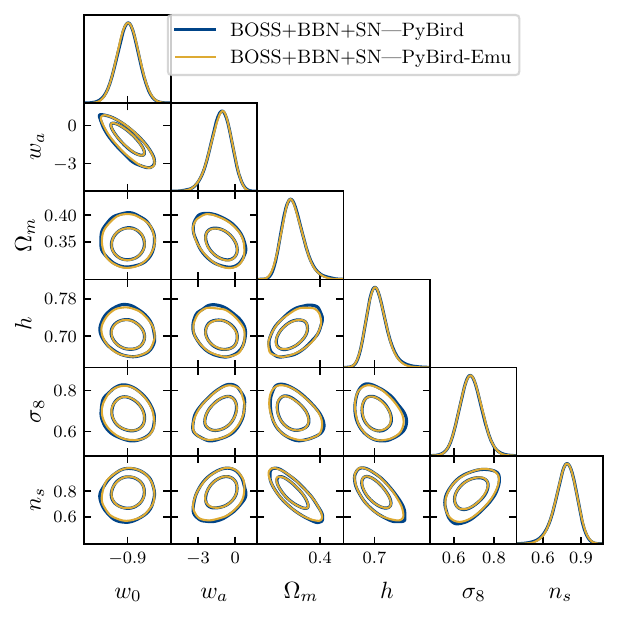}}
  \caption{\textbf{Comparison (I) \PyBird vs. \JAXEmu on BOSS data} --- 1D and 2D marginal posterior distributions of inferred parameters in $\Lambda$CDM (\textit{left panel}) or $w_0w_a$CDM (\textit{right panel}) from BOSS data, with a BBN prior on $\omega_{\rm b}$ (not shown for clarity). 
  The $w_0w_a$CDM fit also includes low-redshift supernova data from Pantheon+. 
  Given the large parameter uncertainties and non-negligible shifts in the means with respect to the central values used for the emulator validation shown in table~\ref{tab:validation_range}, these analysis setups stand as stringent parameter coverage tests for \JAXEmu. 
  The posteriors of \PyBird (\textit{blue contours}) and \JAXEmu (\textit{yellow contours}) are practically indistinguishable for both scenarios, validating the ability of \JAXEmu in recovering cosmological parameters across wide parameter ranges.
  }
  \label{fig:m1}
\end{figure}

\paragraph{$w_0w_a$CDM}
As a first step beyond $\Lambda$CDM, we consider a two-parameter extensions $w_0w_a$CDM. 
Since this model is included in the cosmology bank informing the $\log P_{\rm lin}$ prior of the emulator training set, this serves primarily as a sanity check rather than a challenging test. 
BOSS alone covers only a limited range of redshifts.
In order to get meaningful constraints in $w_0w_a$CDM, we combine the BOSS likelihood with low-redshift data from Pantheon+ supernovae, with their absolute magnitude left uncalibrated. 
Results of the comparison are shown alongside $\Lambda$CDM ones in figs.~\ref{fig:lcdm}~and~\ref{fig:w0wa}, where the posteriors from \PyBird and \JAXEmu are virtually indistinguishable, validating the accuracy of \JAXEmu over a wide range of parameters, given the wide size of the uncertainties. 

\paragraph{}
We now turn to two additional models: early dark energy (EDE)~\cite{1811.04083} and self-interacting neutrinos (SI$\nu$). 
These scenarios were not included in the design of the emulator input coverage, and thus it is not guaranteed that the linear power spectra preferred by the data for these models lie within the emulator's training distribution. 
These fits therefore test the robustness of our emulator design described in sec.~\ref{sec:emulation}, which aims to ensure broad and general coverage in $\log P_{\rm lin}$ space, avoiding out-of-distribution failures.

\begin{figure}[!ht]
  \centering
  \subcaptionbox{EDE\label{fig:ede}}
    [.49\linewidth]{\includegraphics[width=\linewidth]{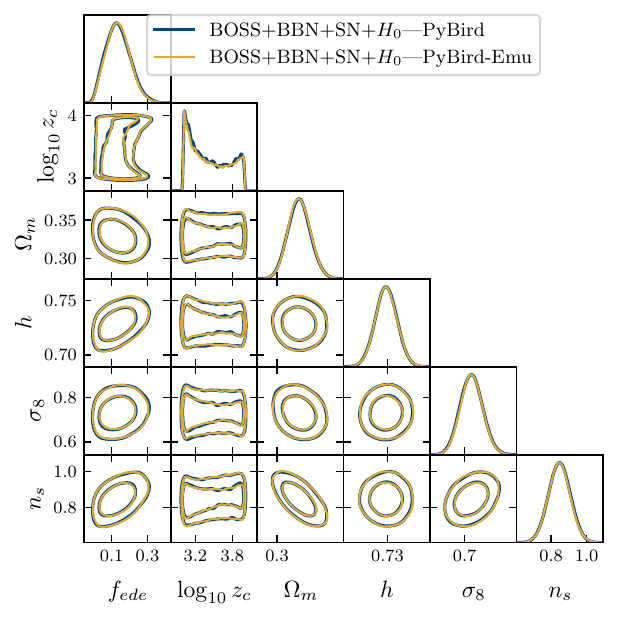}}\hfill
\subcaptionbox{SI$\nu$\label{fig:Geff}}[.49\linewidth]{\includegraphics[width=\linewidth]{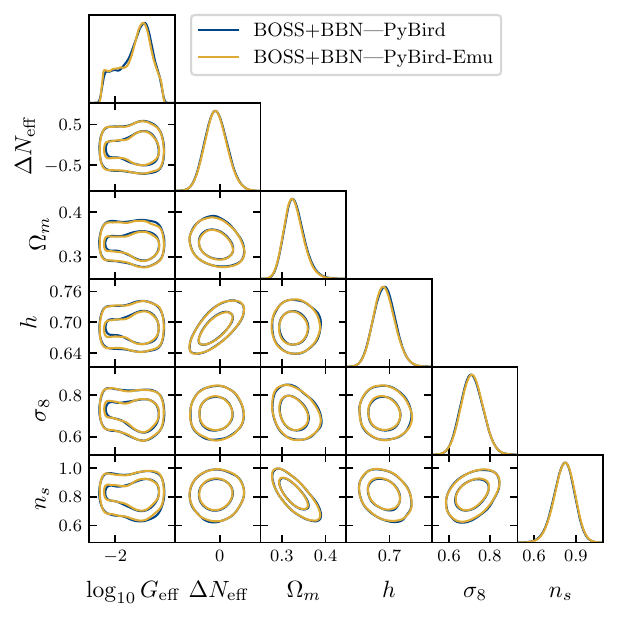}}
  \caption{\textbf{Comparison (II) \PyBird vs. \JAXEmu on BOSS data} --- 1D and 2D marginal posterior distributions of inferred parameters in EDE (\textit{left panel}) or SI$\nu$ (\textit{right panel}) from BOSS data, with a BBN prior on $\omega_{\rm b}$ (not shown for clarity). 
  The EDE fit also includes low-redshift supernova data from Pantheon+ with magnitude calibrated from the S$H_0$ES distance ladder, implying a large $H_0$ compared to its preferred value in $\Lambda$CDM (without S$H_0$ES). 
  We restrict $\log_{10}G_{\rm} \in [-2.5, -0.3]$ to probe exclusively the preferred strongly-interacting regime by BOSS. 
  These analysis setups thus provide far-away cosmologies to test the accuracy of \JAXEmu. 
  Crucially, EDE and SI$\nu$ were not used when designing the emulator's input coverage. 
  The posteriors of \PyBird (\textit{blue contours}) and \JAXEmu (\textit{yellow contours}) are practically indistinguishable for all scenarios, 
  validating the flexibility of \JAXEmu in recovering cosmological parameters across model space. 
  }
  \label{fig:m1}
\end{figure}

\paragraph{Early Dark Energy}
Designed to solve the Hubble tension (see \textit{e.g.},~ref.~\cite{Poulin:2023lkg}), EDE introduces a new component that behaves like dark energy around the time of recombination, that subsequently decays, allowing the universe to follow the standard thermal history. 
The injection of EDE increases the physical sound horizon seen in the cosmic microwave background (CMB), thereby reducing the comoving distance from us to the CMB, and thus increasing $H_0$ accordingly, in order to keep the well-measured angular acoustic scale constant (see \textit{e.g.}, ref.~\cite{Poulin:2024ken} for refined details on subleading effects to account when modifying the sound horizon as early-time solution attempts to the Hubble tension). 
Considering axion-like EDE~\cite{Poulin:2023lkg}, three extra parameters are introduced with respect to $\Lambda$CDM: $z_c$, the redshift at which EDE is injected, $f_{ede}$, the fraction of EDE injected at $z_c$, and the initial axion-like pseudo-scalar field value $\phi_i$ (fixing $n=3$). 
Notably, when fit in combination with S$H_0$ES and Pantheon+~\cite{2112.04510,2202.04077}, a large value of $H_0$ is preferred in this model compared to the preferred value in $\Lambda$CDM by BOSS~\cite{Simon:2022adh}, together with a non-zero $f_{ede}$ at $\sim 2\sigma$. 
Whether EDE remains a viable solutions to the $H_0$ tension is currently under debate, especially in light of new cosmological data (see \textit{e.g.},~refs.~\cite{DAmico:2020ods,Ivanov:2020ril,Smith:2020rxx,Simon:2022adh,Reeves:2022aoi,Efstathiou:2023fbn}). 
As far as we are concerned here, EDE provides a non-trivial example of the cosmology-independence of our emulator. 
For this test, we fit the BOSS data together with Pantheon+ supernovae with magnitude calibrated with the S$H_0$ES distance ladder, and a prior from BBN experiments. 
For this analysis, we use \texttt{AxiCLASS}\footnote{\url{https://github.com/PoulinV/AxiCLASS}}~\cite{Poulin:2018dzj,Smith:2019ihp}, a modified version of \texttt{CLASS}. 
The comparison between results from \PyBird and \JAXBird is shown in fig.~\ref{fig:ede}, where no visible shift between the posteriors can be seen. 

\paragraph{Self-interacting neutrinos}
The other extension we consider is a scenario in which one or more of the three left-handed neutrinos display a non-standard self-interactions (see \textit{e.g.}, ref.~\cite{2203.01955} for a review). 
In presence of self-interactions, the onset of their free-streaming is delayed with respect to their decoupling with the Standard Model bath (around temperature $T\sim 1$ M$e$V), thereby suppressing the anisotropic stress sourcing the difference of the metric potentials. 
Such effect is partially degenerate with the presence of additional relativistic species, parametrised by $\Delta N_{\rm eff}$. 
Intriguingly, hints of strongly interacting neutrinos ($\log_{10} (G_{\rm eff}/\textrm{M}e\textrm{V}^{-2}) \sim -1.5$) have been found in the BOSS data~\cite{2309.03941,2309.03956}, that however vanish once combined with CMB~\cite{Camarena:2024zck}. 
Whether the so-called strongly-interacting regime remains a viable description of the cosmological data does not matter for our concern here. 
Instead, we use the opportunity to test our emulator on far-away cosmologies explicitly not represented when building the training set of our emulator. 
For this test, we fit the BOSS data together with a BBN prior, and use a flat prior on $\log_{10}(G_{\rm eff}/\textrm{M}e\textrm{V}^{-2})$ of $[ -2.5, -0.3]$ to probe the preferred strongly-interacting mode~\cite{2309.03941} (as corresponding to far-away cosmologies useful to test our emulator accuracy). 
We also vary the number of extra relativistic species $\Delta N_{\rm eff}$, and fix the neutrino mass to minimal, $m_\nu = 0.06 \, e$V.  
For this analysis, we use \texttt{class\_interacting\_neutrinos}\footnote{\url{https://github.com/PoulinV/class_interacting_neutrinos}}, a modified version of \texttt{CLASS}. 
Again, no shift are visible between the posteriors obtained with \PyBird and \JAXEmu shown in fig.~\ref{fig:Geff} for this model, thus validating the accuracy of \JAXEmu to recover parameters across a wide range of cosmologies. 

%%%%%

\subsection{Stage-4 LSS survey forecast}\label{sec:sampler}

\begin{figure}[!ht]
    \centering
    \includegraphics[width=0.99\linewidth]{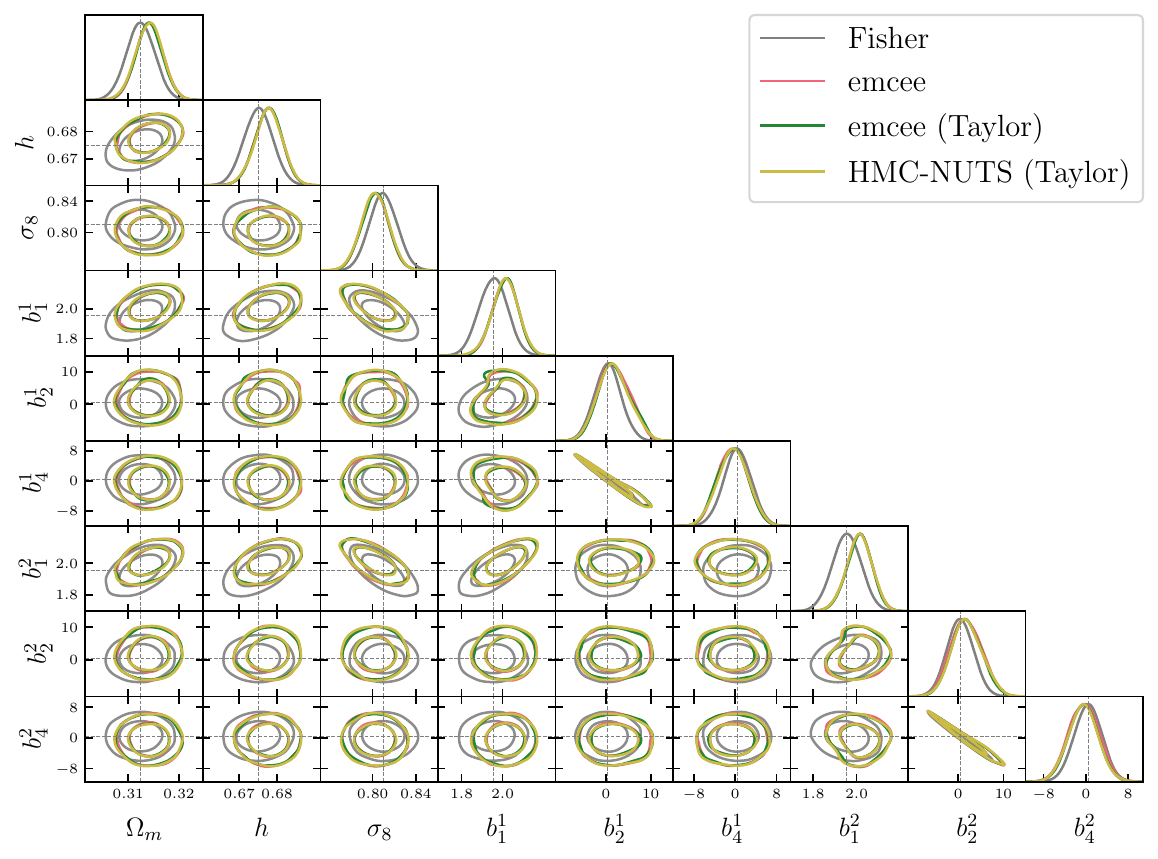}
    \caption{\textbf{Sampler accuracy with \JAXBird for Stage-4 LSS survey forecast} --- 
    1D and 2D marginal posterior distributions of inferred $\Lambda$CDM parameters (with $\omega_{\rm b}$ and $n_s$ fixed) and some representative EFT parameters from DESI Year 6 mock data, obtained using various sampling strategies: \texttt{emcee} with \JAXBird (\textit{red contours}),  \texttt{emcee} with \Taylor (\textit{green contours}), and \texttt{HMC-NUTS} with \Taylor (\textit{yellow contours}). 
    All results agree within convergence. 
    The Fisher forecast (\textit{grey contours}) and the fiducial values (\textit{dashed lines}) are shown for comparison. }
    \label{fig:desi}
\end{figure}

\begin{table}[!ht]
    \centering
    \begin{tabular}{lcc}
    \toprule
    Sampler & ESS/s & Acceptance rate \\
    \midrule
    \texttt{emcee} sequential      & 1.3 & 19\% \\
    \texttt{emcee} vectorised (batch 48) & 3.4 & 19\% \\
    \texttt{HMC-NUTS}              & 2.2 & 90\% \\
    \bottomrule
    \end{tabular}
    \caption{\textbf{Sampler performance with \JAXBird for Stage-4 LSS survey forecast} --- 
    Effective sample size per seconds (ESS/s) and acceptance rate of MCMC chains from ensemble sampling with \texttt{emcee} in sequential or vectorised mode, or gradient-based sampling with \texttt{HMC-NUTS}. 
    These performances are achieved on one GPU core using \Taylor. 
    For this modest 24-dimensional, near-Gaussian posterior, \texttt{emcee} achieves a decent acceptance rate and delivers the best performance when used with vectorised \JAXBird. Although \texttt{HMC-NUTS} attains a significantly higher acceptance, performance gain is limited in this low-dimensional setting by the additional computational cost of gradient evaluations, which are approximately three times more expensive per likelihood call.
    }
    \label{tab:ess}
\end{table}

In this section, we present a forecast for a Stage-4 LSS survey, using specifications inspired by DESI Year 6, extrapolated from ref.~\cite{DESI:2024aax,2503.14738}. 
Details on the generation of the synthetic mock data are provided in our companion paper~\cite{paper2}, with survey characteristics summarised in table 2 therein.
The forecast includes 12 EFT parameters per redshift bin across 7 bins, resulting in a total of 84 nuisance parameters. 
Of these, 9 out of 12 EFT parameters enter the model linearly and are analytically marginalised following appendix D of ref.~\cite{paper2}, leaving 21 nuisance parameters to be sampled numerically in addition to the cosmological parameters.
We focus on $\Lambda$CDM , fixing $\omega_{\rm b}$ and $n_s$ to their fiducial values for simplicity. 
The likelihood, model, and priors are the same as those described at the beginning of sec.~\ref{sec:applications}.
For this forecast, we use \JAXEmu together with our modified version of \texttt{Symbolic-Pk} Boltzmann emulator~\cite{2311.15865,2410.14623} to enable a fast, fully differentiable pipeline. 
For reference, a single likelihood evaluation using \JAXEmu takes approximately 1 ms on a GPU core, with gradient evaluations taking about three times longer. 

While previous analyses relied on the Metropolis-Hastings algorithm implemented in \texttt{MontePython}~\cite{Audren:2012wb,1804.07261}, here we take the opportunity to showcase alternative samplers that leverage the new features of \JAXBird. These include the Fisher matrix and the gradient-based sampler \texttt{NUTS-HMC}~\cite{Hoffman:2011ukg}, both making use of AD, and the ensemble sampler \texttt{emcee}~\cite{1202.3665}, benefiting from vectorisation.
We also compare results obtained using the Taylor expansion of the model, with derivatives computed via AD. 
For details on these \JAX-based features, see sections~\ref{sec:jax}~and~\ref{sec:AD}.
Fig.~\ref{fig:desi} compares the cosmological posteriors obtained with \JAXEmu across the different sampling strategies. Table~\ref{tab:ess} presents the effective sample size per second (ESS/s) for each sampler under various analysis configurations. 

\paragraph{Taylor expansion of LSS observables}
The posteriors obtained using the Taylor expansion of the model---computed to third order in derivatives and referred to as \Taylor---are in excellent agreement with those obtained without the expansion. 
The same holds true for BOSS data (not shown).
Remarkably, \JAXEmu runs in approximately the same time as \Taylor, being only about 0–40\% slower depending on the input batch size. The speed advantage diminishes rapidly as the batch size increases, with virtually no gain observed for batches of 10 inputs or more (on the hardware used in this test).
In more realistic analysis settings that include observational effects, \Taylor may offer greater speed-up. 
It is worth noting, however, that computing the derivatives for \Taylor requires a non-negligible pre-computation time, which is about 150 seconds for our DESI mock survey, amounting to roughly 10\% of the total sampling walltime. 
Overall, it represents a valuable alternative for fast and accurate inference, although we emphasise the importance of validating its accuracy in beyond-$\Lambda$CDM scenarios.

\paragraph{Sampler comparison}  
We compare three posterior sampling strategies: sequential and vectorised ensemble sampling using \texttt{emcee}, and gradient-based sampling with \texttt{HMC-NUTS}. To monitor convergence, we use the integrated autocorrelation time $\tau$, requiring that the number of samples satisfy \( N_{\text{samples}} \gtrsim 50\,\tau \). The effective sample size (ESS) is computed using \texttt{GetDist}~\cite{1910.13970} as
\begin{equation}
\text{ESS} = \frac{N_{\text{samples}}}{2\tau} \ .
\end{equation}
We then compute the ESS per second (ESS/s) by dividing by the total sampling walltime. Importantly, we exclude the burn-in period for \texttt{emcee} and the warm-up phase for \texttt{HMC-NUTS}, during which hyperparameters are tuned. 
Both take approximately an extra timing of about $30\%$ of the sampling one. 
As shown in table~\ref{tab:ess}, \texttt{HMC-NUTS} outperforms sequential \texttt{emcee} in our DESI mock analysis by a factor of 1.7. This improvement is largely due to its higher acceptance rate (90\% for \texttt{HMC-NUTS} vs.\ 20\% for \texttt{emcee}), though tempered by the additional computational cost of gradient evaluations (roughly 3$\times$ slower per likelihood call). The expected speed-up of about 1.5 is consistent with the observed factor of 1.7.
Vectorised \texttt{emcee}, using 48 walkers in parallel, achieves the highest ESS/s of 3.4. This gain aligns with the $\sim$3$\times$ speed-up in model evaluation enabled by processing input batches via vectorised \JAXEmu. 
Based on this specific case study, we conclude that vectorised ensemble sampling is the most efficient strategy for posterior estimation with \JAXEmu, as long as the acceptance rate stays above $\sim 10\%$, assuming that the acceptance rate for \texttt{HMC-NUTS} stays at $\sim 90\%$. 
Other samplers available in \JAXBird and described in sec.~\ref{sec:interface} were excluded from this comparison, as we do not expect them to offer additional insights beyond the representative strategies analysed here. 
For example, while \texttt{MC-HMC} may achieve a higher acceptance rate than \texttt{HMC-NUTS}, it is unlikely to yield a significant overall speed-up --- particularly given potentially long warm-up time required for hyperparameter tuning.
In practice, requiring \( N_{\text{samples}} \gtrsim 50\,\tau \) for each parameter, the representative sampling walltime for our DESI Year 6 forecast ranges from about 7 to 20 seconds (on our GPU hardware), depending on the sampling strategy. 
Therefore, in practice, as long as the acceptance rate remains reasonable, any sampler will converge in a decent amount of time for fiducial LSS analyses; pre-computing time can therefore not be neglected when choosing the best sampling strategy.

\section{Conclusions}\label{sec:conclusions}

We have presented \JAXBird and its model--independent, NN-based emulator \JAXEmu as a new, fully differentiable pipeline for LSS analyses.  
By rewriting the \PyBird EFTofLSS implementation in \texttt{JAX} and emulating the costly loop and IR–resummation terms with NNs, we obtain \emph{order‑of‑magnitude} speed‑ups: one‑loop galaxy power‑spectrum multipoles can be evaluated in $\mathcal{O}(\mathrm{ms})$ on a CPU and $\mathcal{O}(0.1\,\mathrm{ms})$ on a single GPU, a gain of $10^{3}$–$10^{4}$ relative to the original code.  

Crucially, \JAXEmu is \emph{cosmology‑independent}.  
Instead of learning in the space of cosmological parameters, the emulator is trained on the spline coefficients of the linear matter power spectrum on a pre-optimised fixed grid in $k$s. 
To ensure some levels of smoothness while retaining flexibility, we use a Gaussian copula distribution generalising an initial extended cosmological bank of linear power spectrum. 
This design delivers wide coverage and negligible bias: we have validated that \JAXEmu recovers unbiased posteriors, consistent with brute-force \PyBird, even for challenging cases like Early Dark Energy or self-interacting neutrinos that were not explicitly represented in the training set. 
It would be valuable to explore if other decomposition bases and NN architectures are well suited for this problem, eventually achieving higher accuracy and/or flexibility. 
We stress, however, that \JAXEmu is valid to explore cosmologies for which the current EFTofLSS implementation of \PyBird itself is a valid description: Galilean-invariant single adiabatic clock with exact-time dependence.  
For models modifying the structure of the EFTofLSS kernels nontrivially (see \textit{e.g.},~refs.~\cite{Verdiani:2025jcf}), \JAXEmu would need to be trained again.

Because \JAXBird is written entirely in \texttt{JAX}, users inherit AD, JIT compilation, and vectorisation \emph{for free}.  
Machine‑precision gradients and Hessians  with AD enable HMC, Fisher forecasts, and gradient‑based optimisers, while \texttt{vmap} allows hundreds of MCMC walkers to be evaluated in parallel. 
Native GPU support lets the same \Python code scale effortlessly with emerging hardware. 
As a first non-trivial application of \JAXBird, we demonstrate that volume projection effects are mitigated in LSS analyses once defining our expectation values with respect to non-flat measures computed with the power of AD in our companion paper~\cite{paper2}.

These features position \JAXBird as a ready‑made engine for Stage‑4 LSS surveys such as DESI and \emph{Euclid}, where likelihoods will involve $\mathcal{O}(10^{2})$ nuisance parameters and billions of modes.  
Looking ahead, the same strategy can accelerate higher‑order perturbative predictions and complementary observables (\textit{e.g.},~one-loop bispectrum~\cite{DAmico:2022ukl,DAmico:2022osl}, configuration‑space correlation functions~\cite{Zhang:2021yna}, projected angular functions~\cite{Reymond:2025ixl}, multi-tracer spectra~\cite{Mergulhao:2021kip,Mergulhao:2023zso}).   
In summary, together with joint, parallel efforts such as \texttt{effort.jl}~\cite{Bonici:2025ltp}, \JAXBird brings the EFTofLSS into the era of high‑performance, differentiable computing. It couples the flexibility of \Python to the throughput of GPUs, turning previously prohibitive analyses into routine tasks and ensuring that the community can fully exploit forthcoming LSS data.

%%%%%%%%%%%%%%%%%%%%%

\section*{Acknowledgements}
We thank Marco Bonici, Arnaud de Mattia, Martin Kärcher, and Alexandre Refregier for discussions. 
We thank Takahiro Nishimichi for providing the data of the PT Challenge simulations. 
PZ acknowledges support from Fondazione Cariplo under the grant No 2023-1205. 
Some computations were performed on the \textit{Euler} high-performance computing cluster at ETH Z\"urich, and we acknowledge assistance from the support team.

%%%%%%%%%%%%%
%
%
%
%

\appendix

\section{Vectorised FFTLog-based IR-resummation}\label{app:resum}

In this appendix, we review the IR-resummation in \PyBird~\cite{DAmico:2020kxu}, highlighting a key difference in its implementation with its first release: the vectorisation in one single 1D array of the terms appearing in the nested sums of eq.~\eqref{eq:resum_master}. 
Compared to naive nested \texttt{for} loops, this results in a factor $\sim \mathcal{O}(10)$ acceleration for the evaluation of the IR-resummation, which now takes $\sim 1/10$ of the loop evaluation time. 
Overall, \PyBird now takes about $500 \, \textrm{ms}$ (with $\sim 60 \, \textrm{ms}$ spent on the IR-resummation) to evaluate one power spectrum as detailed in sec.~\ref{sec:performance}.  

Contrary to most alternative codes (with the exception of \texttt{Velocileptors}~\cite{Chen:2020fxs}), \PyBird implement the full Lagrangian IR-resummation first derived in refs.~\cite{Porto:2013qua,Senatore:2014via} and further developed in refs.~\cite{Lewandowski:2015ziq,DAmico:2020kxu,2111.05739}, and thus does not rely on the wiggle-no-wiggle split of the linear power spectrum. 
The $j$-th loop term of the multipoles of the IR-resummed power spectrum at the $N$-loop order, $P_\ell^j(k)|_N$, can be written as an integral over the unresummed correlation function multipoles $\xi^j_{\ell'}(s)$ modulated by a convolution matrix $Q_{\ell\ell'}||_{N-j}(k,s)$ (see \textit{e.g.},~\cite{2111.05739}), 
\begin{equation}
P^j_\ell(k)|_N = \sum_{\ell'}^{N_\ell} 4\pi (-i)^{\ell'}\int ds \, s^2 \, Q_{\ell\ell'}||_{N-j}(k,s) \, \xi^j_{\ell'}(s) \ .
\end{equation}
We remind that \PyBird computes swiftly and stably the correlation function analytically using the FFTLog~\cite{DAmico:2020kxu,Zhang:2021yna}. 
Here, $Q_{\ell\ell'}||_{N-j}(k,s)$ represents an exponential damping $F||_{N-j}$ resumming the large, $\mathcal{O}(1)$-long-wavelength displacements, weighted by an exponential phase factor from the Fourier transform together with Legendre polynomials $\mathcal{L_\ell}(\hat k \cdot \hat n) \mathcal{L_{\ell'}}(\hat s \cdot \hat n)$ from the multipole expansions of the redshift-space power spectrum and correlation function (where $\hat n$ is the line-of-sight direction), angle-averaged in both Fourier and configuration space, 
\begin{equation}\label{eq:Q}
Q_{\ell\ell'}||_{N-j}(k,s) = \frac{2\ell+1}{2} \int_{-1}^{1} d (\hat k \cdot \hat n) \int d^2 \hat s \, e^{i ks (\hat k \cdot \hat s)} F||_{N-j}(k, s, \hat k \cdot \hat s) \, \mathcal{L_\ell}(\hat k \cdot \hat n) \mathcal{L_{\ell'}}(\hat s \cdot \hat n) \ .
\end{equation}
At one-loop order, the linear and one-loop terms are resummed using respectively $F||_{1}$ and $F||_{0}$, defined as (see \textit{e.g.},~\cite{2111.05739}) 
\begin{equation}\label{eq:exp_damping}
F||_{0} = \exp\left[-\frac{1}{2}\braket{(\pmb{k} \cdot \Delta\pmb{\psi})^2} \right] \ , \qquad F||_{1} = F||_{0} \times \left[ 1-\frac{1}{2} \braket{(\pmb{k} \cdot \Delta\pmb{\psi})^2} \right] \ . 
\end{equation}
Here, from the two-point function of the linear displacements $\pmb{\psi}$, we have
\begin{equation}
\braket{(\pmb{k} \cdot \Delta\pmb{\psi})^2} = k_ik_j(X_0(s)\delta_{ij} + X_2(s)\hat s_i \hat s_j) \ ,
\end{equation}
where the IR-filters are defined as
\begin{equation}\label{eq:X}
X_0(s) = \frac{1}{3} \int^{\Lambda_{\rm IR}} \frac{dp}{2\pi^2} \, P_{\rm lin}(p) \left[1 - j_0(ps) - j_2(ps)\right] \, , \qquad X_2(s) = \int^{\Lambda_{\rm IR}} \frac{dp}{2\pi^2} \, P_{\rm lin}(p) j_2(ps) \, . 
\end{equation}

To make the evaluation practical, the trick is to Taylor expand the exponential factors in eq.~\eqref{eq:exp_damping} to a sufficiently high order $M$ ($M \sim 20$ in \PyBird) so that the resulting truncated Taylor series are a good numerical approximation of the full exponential~\cite{DAmico:2020kxu}. 
We can then simplify $Q_{\ell\ell'}||_{N-j}(k,s)$ in eq.~\eqref{eq:Q} following ref.~\cite{Lewandowski:2015ziq}. 
The idea is to expand the exponential phase factor in the Legendre decomposition of a plane wave, absorbing the powers of $\hat k \cdot \hat s$ as derivatives acting on it, such that the angular integrals over $d^2\hat s$ can be performed using the orthogonality of Legendre polynomials. 
The resulting integrals over $d(\hat k \cdot \hat n)$ of two Legendre times powers of $\hat k \cdot \hat n$ can be performed analytically. 
At the end of the day, the IR-resummed power spectrum implemented in \PyBird reads~\cite{DAmico:2020kxu}
\begin{equation}\label{eq:resum_master}
P_\ell(k)|_N = P_\ell(k) + \sum_{j=0}^N \sum_{\ell'}^{N_\ell} \sum_{m=1}^M \sum_{\alpha}^{N_\alpha} 4\pi (-i)^{\ell'} k^{2m} \,\mathcal{Q}_{||N-j}^{\ell \ell',m, \alpha}(f) \, \int ds \, s^ 2 \,  \left[ X(s) \right]^m   \xi^j_{\ell'}(s) \, j_{\alpha}(ks) \, ,
\end{equation}
where $\left[ X\right]^m$ represents either $X_0^m$, or $X_0^{m-1}X_2$, etc.,
and $\mathcal{Q}_{||N-j}^{\ell \ell',m, \alpha}(f)$ are simple numerical factors that depends on the growth rate $f$. 
The IR-corrections that upgrade the non-resummed power spectrum $P_\ell$ to the resummed one $P_\ell|_N$ appearing in eq.~\eqref{eq:resum_master} are spherical Bessel transforms (SBTs) of the correlation function multipoles filtered by $[X]^m$, that can be computed swiftly using the FFTLog~\cite{Hamilton:1999uv}.  
Note that we also use the FFTLog to compute $X_0$ and $X_2$ defined in eq.~\eqref{eq:X}, making our overall implementation of the IR-resummation stable numerically and fast. 
To make the algorithm efficient, instead of looping over the sums appearing in eq.~\eqref{eq:resum_master}, we perform a unique SBT\footnote{Implemented at \url{https://github.com/pierrexyz/fftlog}} on a multidimensional array of size $(N_\ell, N_{\rm lin} + N_{\rm loop}, N_\ell, M, N_\alpha)$, where $N_{\rm lin}$ and $N_{\rm loop}$ represent the total number of EFT parameter-independent linear and loop terms, respectively. 
This is much faster than performing $N_\ell \times (N_{\rm lin} + N_{\rm loop}) \times N_\ell \times M \times N_\alpha$ sequential SBTs, as it takes advantages of \texttt{BLAS} routines and avoids slow \texttt{Python} \texttt{for} loops. 
The multidimensional array is then contracted along the relevant axis with $\mathcal{Q}_{||N-j}^{\ell \ell',m, \alpha}(f)$, and finally the IR-correction terms are added to each (unresummed) linear and loop contributions. 
Note that as being fully vectorised, \PyBird implementation is straightforwardly ported to \JAX. 

\begin{figure}[!ht]
    \centering
\includegraphics[width=0.54\textwidth]{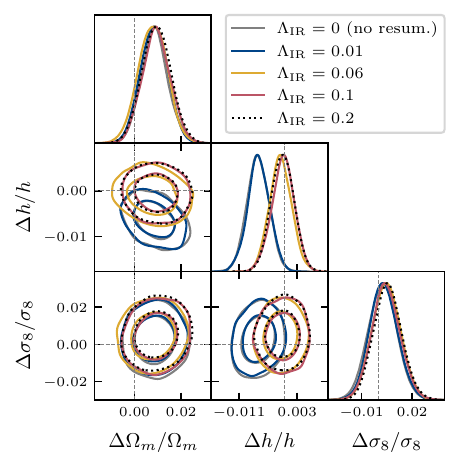}
    \caption{\textbf{$\Lambda_{\rm IR}$-sensitivity on PT challenge data} --- 1D and 2D marginal posterior distributions of inferred $\Lambda$CDM parameters from the PT challenge simulations, with fixed $\omega_{\rm b}$ and $n_s$, as functions of the resummation cutoff $\Lambda_{\rm IR}$. 
    In the limit $\Lambda_{\rm IR} \rightarrow 0$, we recover the results using unresummed predicitions, while for $\Lambda_{\rm IR} > 0.06 \, \hinvMpc$, all results display good agreement. 
    All $\lambda_{\rm IR}$s are in $\hinvMpc$. 
    }
    \label{fig:lambdaIR}
\end{figure}

To close this appendix, we take the opportunity to study the sensitivity of inferred cosmological parameters on the IR-cutoff $\Lambda_{\rm IR}$ entering in eq.~\eqref{eq:X}. 
To do so, we use the PT challenge simulations presented in sec.~\ref{sec:PT}, that we fit with various values for $\Lambda_{\rm IR}$ (implemented as a Gaussian damping rather than a sharp cutoff such that $X_0$ and $X_2$ can be computed using the FFTLog) in fig.~\ref{fig:lambdaIR}. 
For $\Lambda_{\rm IR}$ close to $0$, we recover the same results obtained using unresummed predictions, as it should. 
We see that from $\Lambda_{\rm IR} \sim 0.06 \hinvMpc$ and above, the posteriors of the inferred cosmological parameters remains fairly consistent with each others. 
Based on this comparison, we set $\Lambda_{\rm IR}= 0.1 \hinvMpc$ by default in \PyBird. 

\section{Emulator accuracy up to $k_{\mathrm{max}}=0.3 \, \hinvMpc$}\label{app:emuacc}
\begin{figure}[!ht]
    \centering
    \includegraphics[width=0.99\textwidth]{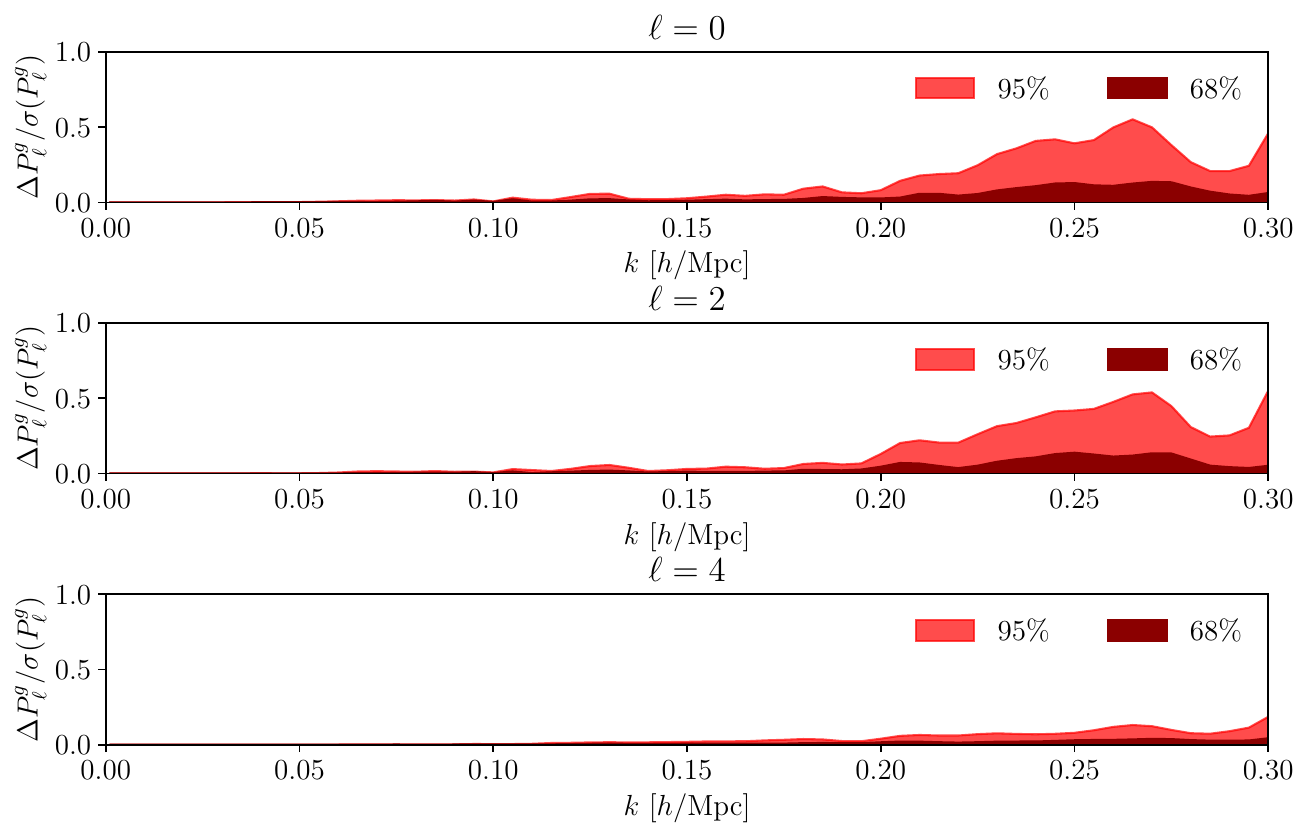}
    \caption{68\% and 95\% quantiles of the differences in the galaxy power spectrum multipoles across scales up to $k_{\rm max} = 0.3 \, \hinvMpc$, computed with \PyBird. 
    The comparison is between the NN-based emulator predictions and the full calculations, evaluated over the independent validation set described in sec.~\ref{sec:performance}. Errors are shown relative to representative uncertainties expected for Stage-4 LSS surveys.}
    \label{fig:emu_higherkmax}
\end{figure}

As mentioned in the main text, our emulators within \PyBird are trained to operate up to a higher $k_{\mathrm{max}}$ than the typical value used in LSS analyses, namely $k_{\mathrm{max}} = 0.2\,\hinvMpc$ (as adopted in our analyses in sec.~\ref{sec:applications}). 
Nonetheless, it is informative to examine emulator performance at higher $k$ values for two main reasons. 
First, the theoretical power spectrum is typically convolved with a window function matrix (see, \textit{e.g.}, ref.~\cite{Beutler:2018vpe}), which has finite support in $k$. 
As a result, the convolved power spectrum at a given $k_*$ receives some contributions from modes with $k > k_*$. 
Second, analyses are sometimes performed in wedges rather than multipoles. In particular, for wedges constructed such that $\mu \sim 0$, the slowly converging series expansion of redshift-space distortions is suppressed. 
This leaves a more rapidly converging series for that wedge, effectively controlled by the EFTofLSS parameters of the real-space expansion, that allow an extended $k$-reach (see sec.~\ref{sec:PT} for further discussions).
In fig.~\ref{fig:emu_higherkmax}, we present the emulator residual errors up to $k_{\mathrm{max}} = 0.3\,\hinvMpc$. 
While the emulator's accuracy degrades somewhat at higher $k$, 95\% of the validation samples remain within $0.5\sigma$ relative to the representative Stage-4 LSS uncertainties, demonstrating that the emulator still performs reasonably well in this extended regime. We note that the volume used for these uncertainties is at least $5$ times larger than the typical per redshift bin volume for DESI.

 \bibliographystyle{JHEP}
 \small
\bibliography{jaxbird}

 \end{document}